\documentclass[sts]{imsart-arxiv_version}

\RequirePackage{amsthm,amsmath,amsfonts,amssymb}
\RequirePackage[authoryear]{natbib}
\RequirePackage[colorlinks,citecolor=blue,urlcolor=blue]{hyperref}
\RequirePackage{graphicx}

\usepackage[colorinlistoftodos]{todonotes}
\usepackage{changes}
\usepackage[ruled,vlined,linesnumbered]{algorithm2e}
\usepackage{comment}
\usepackage{hyperref}

\startlocaldefs
\theoremstyle{plain}

\newtheorem{proposition}{Proposition}

\theoremstyle{remark}

\newtheorem{remark}{Remark}

\newcommand{\metric}{\mathfrak{d}}

\newcommand{\gla}{\,}  
\newcommand{\gra}{}  
\newcommand{\glb}{\left(}  
\newcommand{\grb}{\right)}  
\newcommand{\glc}{\left[}  
\newcommand{\grc}{\right]}  

\newcommand\sect[1]{Section~\ref{#1}}

\newcommand{\expb}[1]{\exp \glb #1 \grb} 
\newcommand{\expc}[1]{\exp \glc #1 \grc} 
\newcommand{\cosb}[2][]{\cos^{#1} \glb #2 \grb}  
\newcommand{\coshb}[2][]{\cosh^{#1} \glb #2 \grb} 

\newcommand{\loga}[2][]{\log^{#1}\! \gla #2 \gra}  
\newcommand{\logc}[2][]{\log^{#1} \glc #2 \grc}  

\newcommand{\R}{\mathbb{R}}

\endlocaldefs

\begin{document}

\begin{frontmatter}
\title{Sampling algorithms in statistical physics: a guide for statistics and machine learning}
\runtitle{Sampling algorithms in statistical physics}

\begin{aug}
\author[A]{\fnms{Michael F.} \snm{Faulkner}\ead[label=e1]{michael.faulkner@bristol.ac.uk}}
\and
\author[B]{\fnms{Samuel} \snm{Livingstone}\ead[label=e2]{samuel.livingstone@ucl.ac.uk}}
\address[A]{HH Wills Physics Laboratory, University of Bristol, UK, \printead{e1}.}
\address[B]{Department of Statistical Science, University College London, UK, \printead{e2}.}
\end{aug}

\begin{abstract}
We discuss several algorithms for sampling from unnormalized probability distributions in statistical physics, but using the language of statistics and machine learning.  We provide a self-contained introduction to some key ideas and concepts of the field, before discussing three well-known problems: phase transitions in the Ising model, the melting transition on a two-dimensional plane and simulation of an all-atom model for liquid water. We review the classical Metropolis, Glauber and molecular dynamics sampling algorithms before discussing several more recent approaches, including cluster algorithms, novel variations of hybrid Monte Carlo and Langevin dynamics and piece-wise deterministic processes such as event chain Monte Carlo. We highlight cross-over with statistics and machine learning throughout and present some results on event chain Monte Carlo and sampling from the Ising model using tools from the statistics literature. We provide a simulation study on the Ising and XY models, with reproducible code freely available online, and following this we discuss several open areas for interaction between the disciplines that have not yet been explored and suggest avenues for doing so.
\end{abstract}


\end{frontmatter}

\section{Introduction}

Sampling algorithms are commonplace in statistics and machine learning -- in particular, in Bayesian computation -- and have been used for decades to enable inference, prediction and model comparison in many different settings.  They are also widely used in statistical physics, where many popular sampling algorithms first originated~\citep{Metropolis1953EquationOfState,Alder1957PhaseTransition,Alder1959StudiesInMolecularDynamicsI,Alder1959StudiesInMolecularDynamicsII}.  At a high level, the goals within each discipline are the same -- to sample from and approximate expectations with respect to some probability distribution -- but the motivations, nomenclature and methods of explanation differ significantly.

Practitioners in Bayesian inference estimate parameter expectations based on fixed hyperparameters 
and input data.  To provide for this, researchers in Bayesian computation typically strive to establish general-purpose sampling algorithms (most notably Markov chain Monte Carlo) and therefore develop theory concerning how a given sampler behaves in a variety of different settings, characterised by features such as how the tails of a distribution decay (e.g. \cite{jarner2000geometric}) or how much the sampler exploits some particular structure of the model (e.g. \cite{papaspiliopoulos2007general}).  The main concern for a given algorithm is often the extent to which it can be widely implemented with little problem-specific tuning.  Different samplers are compared by assessing how performance depends on the dimension of the parameter space (e.g. \cite{roberts2001optimal}) where `performance' is typically defined as either the mixing time or the asymptotic variance of ergodic averages.  Comparisons are usually based on theoretical results, which are complemented with numerical studies to corroborate the theory.

In statistical physics, expectations are studied as \textit{functions} of the hyperparameters (e.g. the temperature) in order to predict the effect of their variation on the physical system of interest.  The primary goal is to describe complex many-particle phenomena in terms of a reduced set of simplified particle--particle interactions, typically using a Boltzmann--Gibbs distribution.  Unlike the Bayesian posterior, these distributions do not depend on input data, but the normalising constant is nonetheless typically intractable.  Physicists compare estimated expectations with experimental data -- both as functions of the relevant hyperparameters -- which leads either to model modifications if there are discrepancies between the results, or to a successful description of the complex phenomena in terms of the simplified set of interactions; the latter may be followed by further predictions that experimentalists then attempt to confirm or refute.  As regards sampling algorithms, there is of course concern for wide applicability, but another and perhaps stronger imperative is to assess sampler performance on a class of important benchmark models.  
Algorithm performance is often defined as the number of computational steps required to generate independent samples, though mixing times are also measured (e.g. \cite{Lei2018IrreversibleMarkovChains}).  Comparisons are made based on the scaling of performance with the number of particles, which is proportional to the dimension of the parameter.

In addition, distributions of interest in statistical physics very often exhibit multi-modality or anisotropy at low temperature, and the ability of an algorithm to navigate this is also a key measure of sampling efficiency.  By contrast, while anisotropy is a common feature of Bayesian posterior distributions -- typically caused by parameter dependencies induced by the data or built into the prior --  multi-modality is usually confined to specific classes of models that have known non-identifiability issues, such as mixture models \citep{jasra2005markov} or neural networks \citep{neal2012bayesian}.  Nevertheless, anisotropy in statistical physics informs Bayesian computation, and indeed Bayesians should also be wary of the critical slowing down that can accompany either feature near the critical temperature.  Examples of multi-modality, anistropy and critical slowing down are given in Section \ref{sec:SimulationsStudies}.


The objective of this work is therefore to review some model problems and sampling algorithms used in statistical physics, but from the perspective of the statistician or machine learner. 
The timing of our contribution is pertinent, as there have been recent parallel advances in nonreversible sampling algorithms in both Bayesian computation and statistical physics.  Statisticians have established much theory assessing the merits of these algorithms (e.g. \cite{bierkens2018high,bierkens2019ergodicity,andrieu2021hypocoercivity,andrieu2021peskun,deligiannidis2021randomized}), while 
physicists have applied them to great effect in many practical scenarios of interest~\citep{Bernard2009EventChain,Bernard2011TwoStepMelting,michel2014generalized,Kapfer2015TwoDimensionalMelting,Kampmann2015MonteCarlo,Michel2015EventChain,Faulkner2018AllAtomComputations,Hoellmer2020JeLLyFysh,Faulkner2022GeneralSymmetryBreaking}.  Interdisciplinary understanding has been at times lacking, however, so that one goal of the present contribution is to support improved communication between these fields -- to aid the cross-pollination of ideas and innovations.

We do not aim to provide an exhaustive review, as this would be impossible within the confines of a single article.  Instead we give a brief overview of statistical physics in \sect{sec:StatisticalPhysics}, before focusing attention on three well-known problems in Section \ref{sec:SomeExampleModels}: phase transitions in the Ising model, the melting transition on a two-dimensional plane and an all-atom model of water.  In \sect{sec:ClassicalSamplingAlgorithms} we discuss three classical sampling algorithms used in statistical physics:  the Metropolis algorithm~\citep{Metropolis1953EquationOfState}, Glauber dynamics~\citep{glauber1963time} and molecular dynamics~\citep{Alder1959StudiesInMolecularDynamicsI,Alder1959StudiesInMolecularDynamicsII}.
In \sect{sec:AdvancedAlgorithms} we review 
some 
more modern alternatives, before presenting some simulation studies in Section~\ref{sec:SimulationsStudies} and a discussion in Section~\ref{sec:Discussion}, in which we suggest open areas for potential collaboration between disciplines. 

\section{Statistical physics}
\label{sec:StatisticalPhysics}

\subsection{Microscopic statistical models}
\label{sec:StatPhysMicroModels}

The fundamental aim of statistical physics is to predict macroscopic physical phenomena using statistical models of microscopic particle--particle interactions.  Physical systems of interest tend to be viewed as collections of particles either restricted to locations on a shared $d$-dimensional lattice (as in Figure \ref{fig:1dIsingConfigurations}) or moving around on a shared compact $d$-dimensional manifold (as in Figure \ref{fig:HardDisks}).  Models of the former are used in \emph{hard condensed matter} to study solid materials, as well as lattice-confined quantum gases~\citep{Roscilde2016FromQuantum} and other similar systems.  Their constituent particles typically remain fixed to each lattice site and interact as a function of their positions and/or some other microscopic quantity, such as their \emph{spin}.  Models of the latter are predominantly used in \emph{soft-matter physics} to study and compare the solid, liquid and gaseous states of a variety of materials, and their constituent particles typically interact as a function of their positions.  Statistical physics is therefore the bridge between microscopic particle--particle interactions and the macroscopic world.

A microscopic statistical model consists of a collection of particles and a set of simplified rules governing their interactions, all of which is encoded in a joint probability distribution for the particle positions or spins.  The \emph{state} of an $N$-particle model encodes the microscopic information and is represented by the parameter $x := (x_1,...,x_N)^T \in \mathcal{M}^N$, where $\mathcal{M}^N$ is the \emph{configuration space} and $\mathcal{M}$ is the \emph{one-particle configuration space}.  In hard condensed matter, $x_i \in \mathcal{M}$ typically describes the spin (or some other microscopic quantity) of particle $i$, $\mathcal{M}$ is often a subset of $\mathbb{Z}$, $\mathbb{R}$ or $\mathbb{R}^2$, and $N = N_1 N_2 \dots N_d$ with $N_i$ the number of lattice sites along the $i^{\rm th}$ dimension of the lattice.  
In soft-matter physics, the picture is somewhat simpler, with $x_i \in \mathcal{M}$ the position (and occasionally the spin) of particle $i$ and $\mathcal{M}$ a compact $d$-dimensional manifold.  
In both cases, the model is then defined by the Boltzmann--Gibbs probability distribution
\begin{equation} \label{eq:bgdist}
\pi(x; \beta, \theta, N) \propto e^{ -\beta U(x; \theta, N) } ,
\end{equation}
where the \emph{inverse temperature} $\beta > 0$ is the reciprocal of the system temperature with units such that $\beta U$ is dimensionless, and the potential energy $U:\mathcal{M}^N \to \mathbb{R}$ encodes the particle--particle interactions, the number of particles $N$ and a vector of additional hyperparameters $\theta$.  In soft matter, the particle density $\eta := \gamma N / V$ is always a component of $\theta$, with $V$ the volume of $\mathcal{M}$ and $\gamma > 0$ a dimensionless constant. 

In addition, physicists view some microscopic model recast in terms of different hyperparameters as the same model but in a different \emph{statistical ensemble}.  For example, the soft-matter model described at the end of the previous paragraph is in the \emph{canonical ensemble} in which $N$, $V$ and $\beta$ are fixed, but it may be re-expressed in some other statistical ensemble, such as the \emph{grand canonical ensemble} in which $V$ and $\beta$ are fixed, but the number of particles can fluctuate with some fixed potential cost.  Thermodynamic theory then provides a bridge between different statistical ensembles, which can be useful when comparing numerical simulations with physical experiments.  In the remainder, we assume $N$, $V$ and $\beta$ are fixed, unless otherwise stated.

\subsection{Periodic boundary conditions and the thermodynamic limit}
\label{sec:PeriodicBoundaries}

To remove boundary effects in hard condensed matter, researchers typically apply \textit{periodic boundary conditions} by choosing the shared $d$-dimensional lattice (on which particle locations are restricted) to have $d$-dimensional toroidal topology: for a lattice of $N_1 \dots N_d$ particle sites, we identify lattice site $(y_1, y_2, \dots, y_d)$ with $(y_1 + N_1, y_2, \dots, y_d) \sim \dots \sim (y_1, y_2, \dots, y_d + N_d)$.  For example, the one-dimensional Ising configuration in Figure \ref{fig:1dIsingConfigurations} is on a ring lattice.  This better reflects the macroscopic systems under consideration, in which boundary effects are usually negligible compared to the large bulk of the system. 
Analysis is performed on  the $N_1 \dots N_d$-site lattice before the \emph{thermodynamic limit} is taken by letting $N_1, \dots , N_d \to \infty$ with the ratio $N_1 : \dots : N_d$ fixed.  

In soft-matter physics, the one-particle configuration space $\mathcal{M}$ is typically chosen to be the $d$-dimensional torus $\mathbb{T}^d$ of volume $V = L^d$, which can be defined as $d$-dimensional Euclidean space under the identification $(x_1, x_2, \dots, x_d) \sim (x_1 + L, x_2, \dots, x_d) \sim \dots \sim (x_1, x_2, \dots, x_d + L)$, where $L > 0$ is the linear size of the torus (note that differing linear sizes can also be chosen along each dimension).  For example, the right and left / top and bottom `boundaries' in Figure \ref{fig:HardDisks} are identified with each other.  More formally, this is the quotient space $\R^d / \mathbb{L}^d$, where $\mathbb{L} := L\mathbb{Z}$.  Again, this better reflects the negligible boundary effects in macroscopic systems.  Analysis is typically performed on $N$ particles on a torus of finite volume $V$, and the thermodynamic limit is then taken by setting $N \propto V$ and letting $N \to \infty$.  
In this paper, the \textit{minimal separation vector} $x_{ij}$ between particles $i$ and $j$ is defined as the shortest vector from $x_j$ to $x_i$ on $\mathbb{T}^d$, and 
\begin{align}
    \metric(x_i, x_j) := \|x_{ij}\|_2
\end{align}
is their \emph{minimal separation distance}.  The metric $\metric$ can also be induced from the quotient space representation of $\mathbb{T}^d$ given above.

\subsection{Observables, phase transitions and fundamental axiom}
\label{sec:PhaseTransitions}

An \emph{observable} of the model is any function of its state and hyperparameters, and well-chosen observables allow for the study of thermodynamic phase space and phase transitions. For an observable $\chi(x; \beta, \theta, N)$ the \emph{expected observable}\footnote{Rather than $\mathbb{E} \glc \cdot  \grc$, physicists tend to represent expectations using the notation $\langle \cdot \rangle$.} will be denoted
\begin{align}
    \check{\chi}(\beta, \theta, N) := \mathbb{E}[\chi(x;\beta,\theta,N)].
\end{align}
The \textit{thermodynamic phase space} of some observable is the space of all possible values of the \emph{thermodynamic observable}
\begin{align}
\lim_{N \rightarrow \infty} \check{\chi}(\beta, \theta, N) .
\label{eq:ThermodynamicObservable}
\end{align}
A \textit{thermodynamic phase} is any open and connected region of thermodynamic phase space in which the thermodynamic observable is analytic in both $\beta$ and $\theta$, and a \emph{phase transition} is any boundary between two distinct thermodynamic phases.  Different thermodynamic phases therefore correspond to strikingly different values of some thermodynamic observable, and distinct thermodynamic phases are separated by one or more non-analytic boundaries, each of which indicates a phase transition.  For example, two thermodynamic observables associated with the two-dimensional Ising model (introduced in Section \ref{sec:Ising}) are presented in Figure \ref{fig:2dIsingSpecHeatAndAbsMagDensityAnalytical}, both of which exhibit a phase transition.  


Typically when studying phase transitions an observable is a sum of $\mathcal{O}(1)$ random variables per particle (e.g. the magnetic density of the Ising model in \eqref{eq:MagneticDensity}).  This leads us to a fundamental axiom of statistical physics.  For any such observable, if the ratio of 
the standard deviation and expectation of its norm 
can be made arbitrarily small with increasing particle number $N$, then there exists some finite particle number at which the expectation is considered to have `reached the thermodynamic limit,' as fluctuations from the thermodynamic value are immeasurably small.  This tends to apply to macroscopic physical systems composed of large numbers of particles, though exceptions do occur near phase transitions and other regions of thermodynamic phase space that exhibit power-law correlations 
(e.g.~\cite{Archambault1997MagneticFluctuations,Faulkner2022GeneralSymmetryBreaking}).  The consequence of this axiom is that simulations based on a large but finite number of particles can approximate behaviour in the thermodynamic limit.


\subsection{Entropy, free energy and equation of state}
\label{sec:Entropy}

The (dimensionless) \emph{entropy} $S(\beta, \theta, N) := -\mathbb{E}[ \loga{\pi(x; \beta, \theta, N)}]$ is well-known to statisticians as a measure of the uncertainty associated with the probability distribution $\pi$.  Boltzmann--Gibbs distributions can exhibit varying degrees of multi-modality or anisotropy depending on the values of hyperparameters such as the temperature, meaning phase transitions are often captured by changes in entropy.  Indeed, if not carefully designed, a sampling algorithm might lose access to certain regions of non-negligible probability mass in some low-entropy thermodynamic phase (on a timescale that diverges with particle number $N$).  This may also reflect a loss of \emph{physical ergodicity}~\citep{Palmer1982} if the dynamics of the sampling algorithm are sufficiently similar to those found in nature.  

The \emph{free energy} $F(\beta, \theta, N) := \check{U}(\beta, \theta, N) - \beta^{-1} S(\beta, \theta, N)$ describes the competition between the expected potential and the entropy.  It provides a toolkit for making predictions about thermodynamic phases because it can be expressed analogously to a marginal log-likelihood function: $F(\beta, \theta, N) = - \beta^{-1} \logc{Z(\beta, \theta, N)}$,
where the \emph{partition function} $Z(\beta, \theta, N) := \int \expc{- \beta U(x; \theta, N)} dx$ is the normalising constant of the Boltzmann--Gibbs distribution.  Free energies at different fixed values of $\theta$ can then be compared to predict the most likely state of matter at any given temperature $\beta^{-1}$.  For example, simulations of a model fluid 
can be performed at some fixed temperature with two different values of the particle density $\eta$ (with the rest of $\theta$ fixed) 
where one is a possible density of the gaseous phase ($\eta_{\rm g}$) and the other is a possible density of the liquid phase ($\eta_{\rm l} > \eta_{\rm g}$).  If the free energy at $\eta_{\rm g}$ is less than the free energy at $\eta_{\rm l}$, it would follow that the fluid is more likely to be in a gaseous state than a liquid state at the chosen temperature.  This is analogous to marginal log-likelihood model comparison in Bayesian inference, where the two fixed values of $\eta$ are equivalent to the two models being compared, and analogies can be drawn between the single fixed temperature and the fixed data of the Bayesian statistical model.

The free energy is also a cumulant generating function, so that it can be used to derive useful expected observables such as the expected (dimensionless) \emph{specific heat} 
\begin{align}
\check{C}(\beta, \theta, N) = - \beta^2 \partial_{\beta}^{2} \glb \beta F(\beta, \theta, N) \grb = \beta^2 {\rm Var} \glc U(x; \theta, N) \grc .
\label{eq:ExpectedSpecificHeat}
\end{align} 
This can be useful when classifying phase transitions.  In addition, the \emph{equation of state} in a soft-matter model is defined via the expected \emph{pressure} 
\begin{align}
\check{p}(\beta, \theta, N) := - \partial_V F \glb \beta, \theta (N, V), N \grb .
\label{eq:EqnOfState}
\end{align} 
This may be familiar to the statistician, as the original work of~\cite{Metropolis1953EquationOfState} entitled `Equation of State Calculations by Fast Computing Machines' applied the Metropolis algorithm to the two-dimensional hard-disk model in an attempt to estimate its equation of state.  Moreover, higher quality estimations of this equation of state were used to identify the fluid--hexatic phase transition described in Section \ref{sec:CoexistenceInterval} \citep{Bernard2011TwoStepMelting}.

\section{Some example models}
\label{sec:SomeExampleModels}

In this section, we present some common models from statistical physics.  In Section \ref{sec:Ising}, we present the \emph{Ising model} and an analysis of its thermodynamic phases. This is a hard-condensed-matter model of magnetism and provides an example of both analytically tractable free energies and \emph{critical slowing down} at a phase transition.  We then comment on the Potts and XY models before devoting the remainder of the section to soft-matter physics.  In Section \ref{sec:harddisk}, we present the \emph{two-dimensional hard-disk model}.  This is possibly the simplest model of particle--particle interactions in soft matter and captures the physics of the \emph{melting transition} in two spatial dimensions.  With this basis, we then move on to the more complex physical interactions used in molecular modelling, where researchers model materials such as pure water or a collection of polymers in a liquid solvent.  The full potential of an \emph{all-atom model} of some molecular system is typically formed by combining several sub-potentials, each of which models a specific force.  In Sections \ref{sec:LennardJones} and \ref{sec:coulomb}, we present (respectively) the \emph{Lennard-Jones potential} and the \emph{Coulomb potential}.  We then introduce two common potentials used to represent molecular-bond bending and stretching in Section \ref{sec:bonded}, before using all four sub-potentials to construct a full all-atom model of water in Section \ref{sec:allatom}. 

\subsection{Ising model}
\label{sec:Ising}

\begin{figure}
\includegraphics[width=0.45\linewidth]{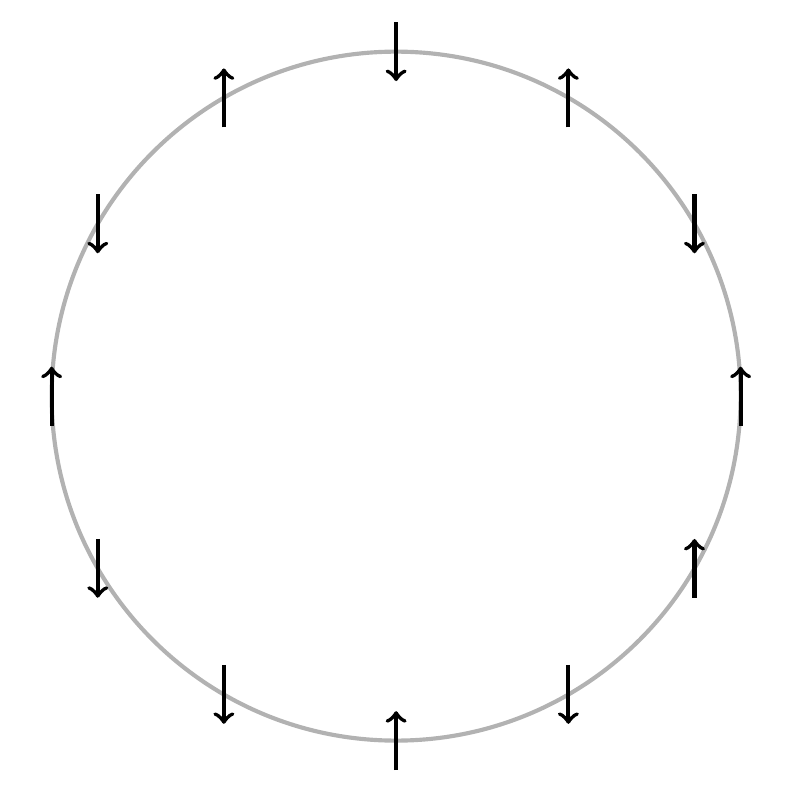}
\caption{Example configuration of the one-dimensional Ising model on a circular ring lattice.  Each up/down arrow represents the spin value $x_i = \pm 1$ of some particle $i$.}
\label{fig:1dIsingConfigurations}
\end{figure}

The $d$-dimensional Ising model is possibly the most well-known model of statistical physics and describes a collection of $N$ particles fixed at the sites $ y_1 , \dots , y_N $ of a regular $d$-dimensional cubic lattice (with toroidal topology).  It was originally constructed as a simple model of $d$-dimensional magnets~\citep{Ising1925Beitrag}, but has since been implemented to model many other physical and non-physical systems.  Most importantly, it is commonly viewed as a paradigmatic model of phase transitions as it is possible to compute its free energy analytically in $d = 1$~\citep{Ising1925Beitrag} and $d = 2$~\citep{Onsager1944CrystalStatistics} dimensions.  Physicists refer to analytical free-energy computation as `exactly solving' the model in question~\citep{Baxter2016ExactlySolved}.

The ferromagnetic Ising model is defined by the potential
\begin{align}
U_{\rm Ising}(x; J, h, N) := - \frac{J}{2} \sum_{i = 1}^N \sum_{j \in S_i} x_i x_j - h \sum_{i = 1}^N x_i  ,
\end{align}
where $J > 0$ is the \emph{exchange constant}, $h \in \mathbb{R}$ controls the strength of an external magnetic field, $S_i$ is the set of the $2d$ neighbours of particle $i$, and $x_i = \pm 1$ is the \emph{spin} of particle $i$.\footnote{The notation $2\sum_{\langle i, j \rangle}$ is often used in place of $\sum_{i = 1}^N \sum_{j \in S_i}$.}  An example configuration in $d = 1$ dimensions is shown in Figure \ref{fig:1dIsingConfigurations}.  The exchange constant $J$ controls the level of correlation between spin values at neighbouring sites. 
Setting $J < 0$ defines the antiferromagnetic Ising model, in which neighbouring spin values are negatively correlated.

\subsubsection{One-dimensional case.}

Ising showed that the one-dimensional Ising model can be solved analytically~\citep{Ising1925Beitrag}.  We show in the supplement \citep{faulkner2023supplement} that the free energy is 
\begin{align}
F_{{\rm Ising, } d = 1}(\beta, J, h, N) = - \beta^{-1} \logc{\lambda_+^N (\beta, J, h) + \lambda_-^N (\beta, J, h)} . 
\label{eq:1dIsingFreeEnergy}
\end{align}
where 
\begin{align}
    \lambda_{\pm} (\beta, J, h) = e^{\beta J} \glc \coshb{\beta h} \pm \sqrt{\sinh^2(\beta h) + e^{-4\beta J}} \grc. \nonumber
\end{align} 
The free energy and all of its derivatives are therefore analytic.  It follows that no thermodynamic observable constructed from derivatives of the free energy exhibits a phase transition.

\subsubsection{Two-dimensional case.}
\label{sec:2DIsing}

\begin{figure}
\includegraphics[width=0.5\linewidth]{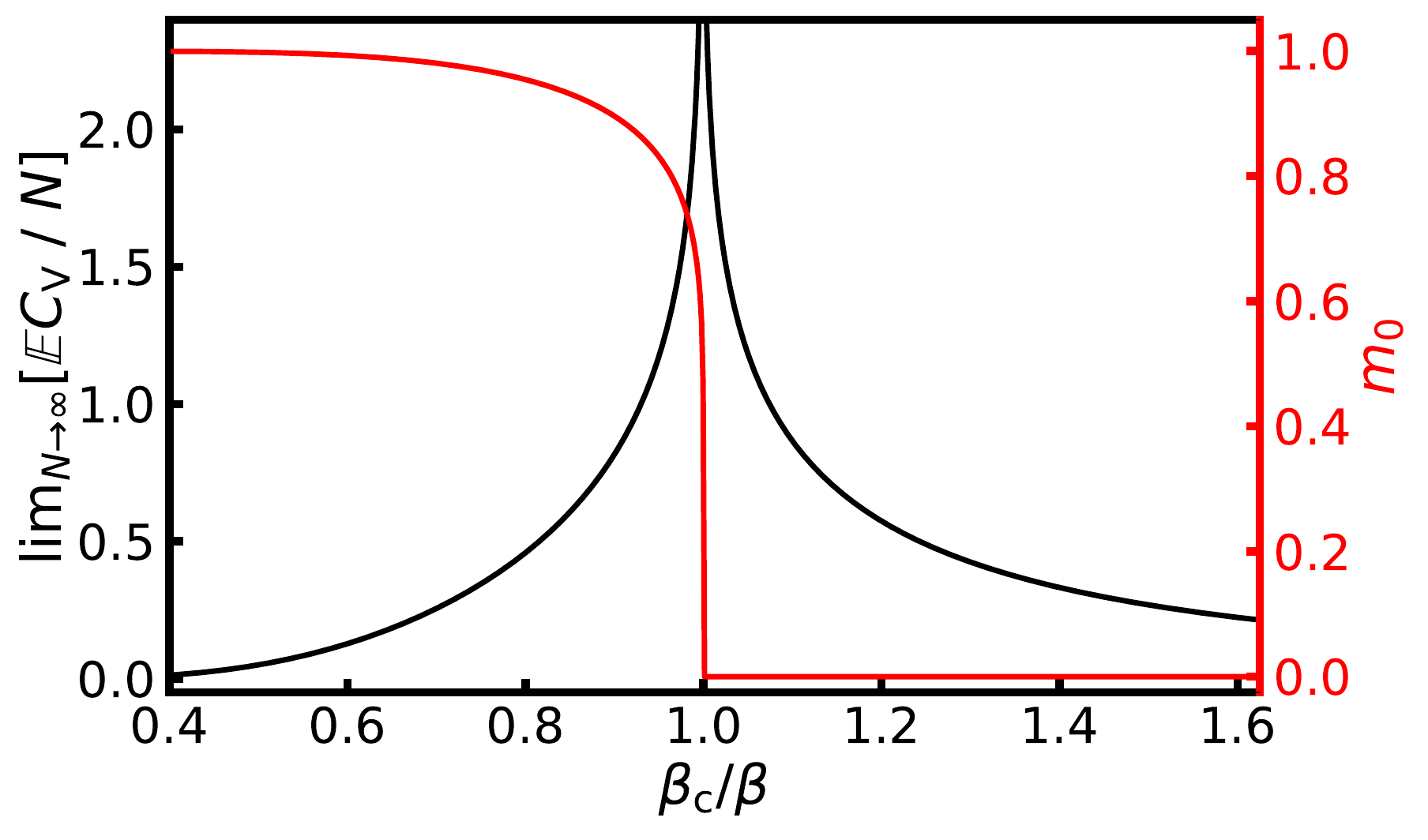}
\caption{Thermodynamic zero-field ($h = 0$) specific heat per particle (black curve; left-hand axis; see \eqref{eq:2dIsingExpectedSpecHeat}) and spontaneous magnetic density (red curve; right-hand axis; see \eqref{eq:2dOnsagerYangSolution}) of the two-dimensional Ising model, both as functions of $\beta_{\rm c} / \beta$.}
\label{fig:2dIsingSpecHeatAndAbsMagDensityAnalytical}
\end{figure}

Building on the initial work of \cite{Kramers1941StatisticsOfTwoDimFerromagnetI,Kramers1941StatisticsOfTwoDimFerromagnetII}, \cite{Onsager1944CrystalStatistics} showed that the two-dimensional Ising model can also be solved analytically.  
The calculations are more involved, but the 
thermodynamic zero-field ($h = 0$) specific heat (see \eqref{eq:ExpectedSpecificHeat}) per particle is 
\begin{align}
    \lim_{N \to \infty} \left( 
    \frac{1}{N} \check{C}_V(\beta, J, h = 0, N) \right) = \beta^2 \partial_{\beta}^2 \gamma(\beta J) ,
\label{eq:2dIsingExpectedSpecHeat}
\end{align}
where 
\begin{align}
\gamma(\beta J) := \ln \left( 2\cosh (2\beta J) \right) + \frac{1}{\pi} \int_0^{\pi / 2} \ln \left[ \frac{1}{2} \glb 1 + \sqrt{1 - \frac{4 \sinh^2 (2\beta J) \sin^2w}{\cosh^4 (2\beta J)}} \grb \right] dw .
\end{align}
It then follows that the thermodynamic 
zero-field specific heat per particle 
diverges logarithmically 
at the \emph{inverse critical temperature} $\beta_{\rm c} := \ln (1 + \sqrt{2}) / (2J)$. 
This predicts a phase transition at $\beta = \beta_{\rm c}, h = 0$, as supported by the black curve in Figure \ref{fig:2dIsingSpecHeatAndAbsMagDensityAnalytical}.  In addition, the \textit{magnetic density} 
\begin{align}
    m(x; \beta, J, h, N) := \frac{1}{N} \sum_i x_i 
    \label{eq:MagneticDensity}
\end{align} 
can also be used to demonstrate the phase transition, 
where 
\cite{Onsager1949Discussion} and \cite{Yang1952SpontaneousMagnetization} proved a non-differentiability in the \emph{spontaneous magnetic density} 
\begin{align}
    m_0(\beta J) := \lim_{h \downarrow 0} \lim_{N \to \infty} \check{m}(\beta, J, h, N) = 
    \begin{cases}
        \left( 1 - (\sinh (2 \beta J))^{-4} \right)^{1 / 8} & {\rm for} \,\, \beta > \beta_{\rm c} , \\
        0 & {\rm for} \,\, \beta < \beta_{\rm c} .
    \end{cases}
\label{eq:2dOnsagerYangSolution}
\end{align}
This provides further evidence of the phase transition at $\beta = \beta_{\rm c}, h = 0$ (as supported by the red curve in Figure \ref{fig:2dIsingSpecHeatAndAbsMagDensityAnalytical}) with the additional insight that it is one between a low-temperature ($\beta > \beta_{\rm c}$) \emph{ferromagnetic} (ordered) phase and a high-temperature ($\beta < \beta_{\rm c}$) \emph{paramagnetic} (disordered) one.  We present a detailed simulation study of this model in Section~\ref{sec:IsingSims}.

\subsubsection{Comments on Potts and XY models.}
\label{sec:XYandPotts}

The $d$-dimensional Potts model~\citep{Potts1952SomeGeneralized} is a generalisation of the $d$-dimensional Ising model, this time with $x_i \in \{ 1, 2, \dots, q \}$ ($q \ge 2$ is an integer) and potential 
\begin{align}
    U_{\rm Potts}(x; J, N) := - \frac{J}{2} \sum_{i = 1}^N \sum_{j \in S_i} \mathbb{I}[x_i=x_j] .
\end{align} 
When $q = 2$, the Potts model is equivalent to the zero-field Ising model.  As well as to phase transitions, the Potts model has been successfully applied to image processing \citep{storath2015joint}.  

The $d$-dimensional XY model can be thought of as another generalisation of the $d$-dimensional Ising model.  Rather than on $\{ -1, +1 \}$, each XY spin is contained in $[0, 2\pi)$, and the XY potential is 
\begin{align}
    U_{\rm XY}(x; J, h, N) := - \frac{J}{2} \sum_{i = 1}^N \sum_{j \in S_i} \cosb{x_i - x_j} - h_{\rm XY} \cdot \sum_i \begin{pmatrix} \cos x_i \\ \sin x_i \end{pmatrix} ,
\end{align} 
where $h_{\rm XY} \in \mathbb{R}^2$.  The $d = 2$ case leads to incredibly rich physics which has been a significant focus of theoretical statistical-physics research since the 1960s~\citep{Salzberg1963EquationOfStateTwoDimensional,Mermin1966AbsenceFerromagnetism,Hohenberg1967ExistenceOfLong-RangeOrder,Berezinskii1973DestructionLongRangeOrder,Kosterlitz1973OrderingMetastability,Kosterlitz1974CriticalProperties,Jose1977Renormalization,Bramwell1993Magnetization,Vallat1994CoulombGas,Archambault1997MagneticFluctuations,Bramwell1998Universality,Jose2013FortyYearsOfBKT,Faulkner2015TSFandErgodicityBreaking,Faulkner2022GeneralSymmetryBreaking}.  We present a simulation study of this model in Section~\ref{sec:XYSims}.

\subsection{Two-dimensional hard-disk model}
\label{sec:harddisk}

\begin{figure*}
\includegraphics[width=0.45\linewidth]{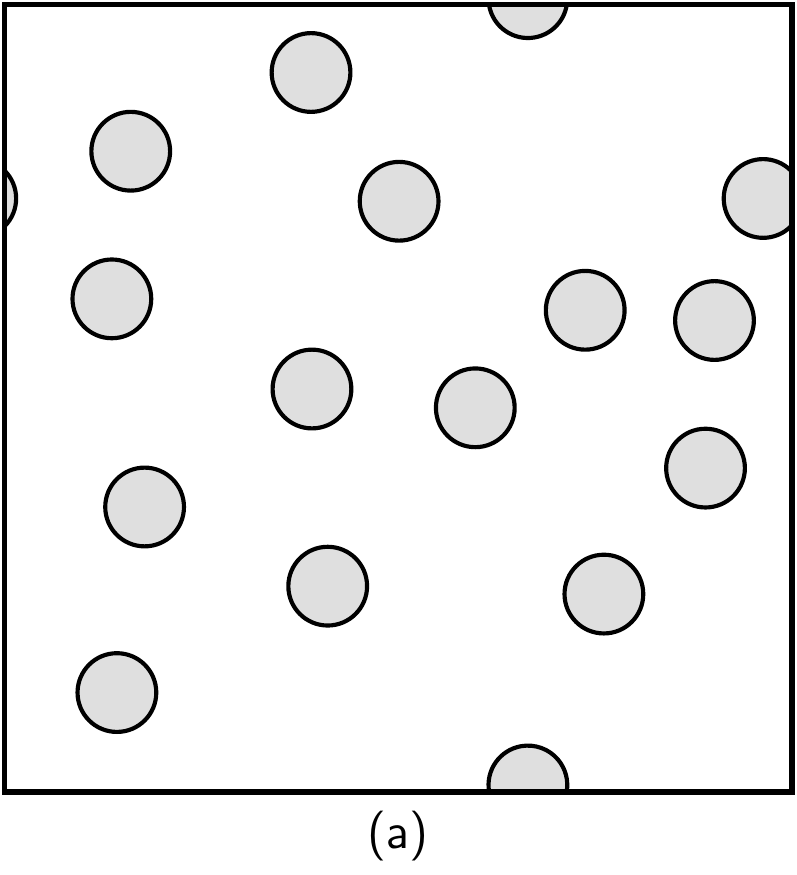}
\hspace{1.0em}
\includegraphics[width=0.45\linewidth]{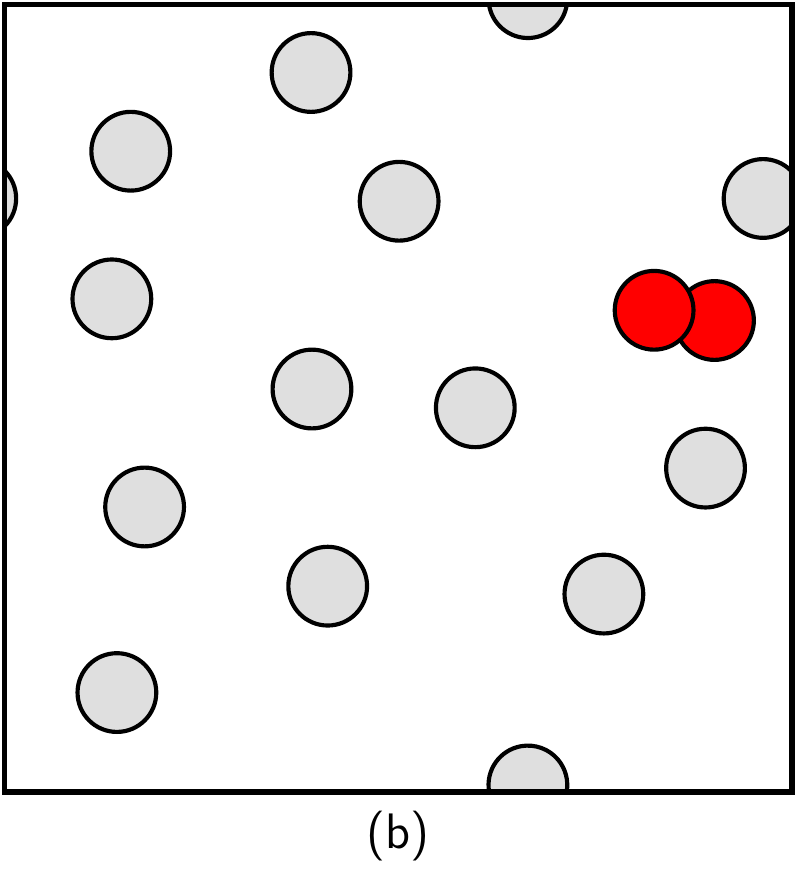}
\caption{Examples of configurations that have non-zero (a) and zero (b) probability density in the two-dimensional hard-disk model.  Two disks pass through the periodic `boundaries' in each example.  The red disks in (b) induce its zero-valued probability density.  These configurations are accepted (a) and rejected (b) by the Metropolis algorithm described in Section~\ref{sec:ClassicalSamplingAlgorithmsMetropolis}.}
\label{fig:HardDisks}
\end{figure*}

The two-dimensional hard-disk model is perhaps the simplest approach to modelling short-range repulsive particle--particle interactions in soft-matter physics.  It is defined by the probability density 
\begin{equation}
\pi(x; \eta, N) \propto \prod_{1 \leq i < j \leq N} \mathbb{I} \glc \metric(x_i, x_j) > 2 \sigma \grc 
\end{equation}
and describes a collection of $N$ identical circular disks of radius $\sigma > 0$, where each exists on the compact manifold $\mathbb{T}^2$ and $\eta = N \pi \sigma^2 / L^2$ is the disk density.  All configurations in which no two disks overlap are equally likely, while all others have zero probability density (examples of two such configurations are shown in Figure \ref{fig:HardDisks}).  The model is therefore independent of the inverse temperature $\beta$ and the sole hyperparameter of interest is the disk density $\eta$.  It can be viewed as the $k \to \infty$ limit of the \emph{two-dimensional soft-disk model}, which is defined on the same parameter space but with the potential $U_{\rm soft-disk}(x; k, \eta, \varepsilon, N) := \sum_{i < j} U_{\rm sd}(x_i, x_j; k, \eta, \varepsilon)$, where 
\begin{equation}
U_{\rm sd}(x_i, x_j; k, \eta, \varepsilon) := \varepsilon \glc \frac{2\sigma}{\metric(x_i, x_j)} \grc^k
\label{eq:SoftDiskModel}
\end{equation}
is the two-particle soft-disk potential, with $\varepsilon > 0$ some constant with units of energy.  The hard-disk model is usually studied on its own, but is also used as a sub-potential in more complex models of attractive particle--particle interactions in which the other sub-potentials contain divergences at $\metric(x_i, x_j) = 0$.

Despite its simplicity, the hard-disk model is able to recreate both fluid and solid structures on a two-dimensional plane, and can therefore be used to investigate the \emph{melting transition} in two spatial dimensions.  In the three-dimensional analogue of hard spheres (each) on $\mathbb{T}^3$, the surface area through which each particle can move is sufficiently large that the effect of movement on other particles diminishes rapidly with $\metric(x_i, x_j)$, so that the three-dimensional model can easily form a solid at densities below the high-density limit.  Conversely, in the one-dimensional analogue, each particle can move only along the single axis of the one-dimensional torus $\mathbb{T}$, so that the effect of movement is significant at all separation distances, and the system cannot form a solid at any density below the high-density limit.  The two-dimensional hard-disk model is an intermediate case whose complete thermodynamic behaviour had eluded researchers for around sixty years~\citep{Metropolis1953EquationOfState,Alder1957PhaseTransition,Kosterlitz1973OrderingMetastability,Halperin1978TheoryTwoDimensionalMelting,Young1979Melting} until event chain Monte Carlo simulations~\citep{Bernard2011TwoStepMelting} led to a theory for its melting transition.  The theory was then corroborated by both molecular dynamics and massively parallel Metropolis simulations~\citep{Engel2013HardDiskEquationOfState} and subsequently confirmed in physical experiments on a collection of colloids on a two-dimensional plane~\citep{Thorneywork2017TwoDimensionalMelting}.  This provides a basis for the melting transition in more complex soft-matter systems, such as films, suspensions and the crossover between two- and three-dimensional behaviour~\citep{Peng2010Melting}.  Here we review the theory.

\subsubsection{Positional correlations and solid phase.}

The \emph{positional correlation function} is
\begin{equation}
g_{\rm p}(x; r, \epsilon, \eta, N) := \sum_{i < j} \mathbb{I} \glc |r - \metric(x_i,x_j)| < \epsilon \grc
\label{eq:PostitionalCorrelations}
\end{equation}
for all $r > 2\sigma$. For a fixed configuration $x$ and a chosen $r$ and $\epsilon>0$, it measures the number of particle pairs whose separation distance is within $\epsilon$ of $r$.  This provides information about the transition into the solid phase.  
Its expected value exhibits a drastic change in behaviour as a function of $r$ at the particle density $\eta = \eta_{\rm s} \simeq 0.720$.  More precisely, Figures 3(b) and S8 of \cite{Bernard2011TwoStepMelting} show that for suitably small $\epsilon$ and $N = 1024^2$
\begin{align}
\check{g}_{\rm p}(r, \epsilon, \eta, N) \propto 
\begin{cases}
\exp \glb - r / \xi_{\rm p}(\eta) \grb \, & \textrm{for} \,\, \eta = 0.718, r > 10 \sigma , \\
r^{- 1 / 3} & \textrm{for} \,\, \eta = 0.720, r > 10 \sigma ,
\end{cases}
\end{align}
where $\xi_{\rm p}(\eta)$ is the \emph{positional correlation length} of the non-solid phases and $x \propto y$ implies that $x = Cy$ for some $C \neq 0$.  This demonstrates exponentially decaying positional correlations for all $r > 10 \sigma$ at $\eta = 0.718$ and power-law decaying positional correlations for all $r > 10 \sigma$ at $\eta = 0.720$.  
These results are consistent with a phase transition from positional disorder to a solid phase with quasi-long-range positional order as $\eta$ increases through $\eta = \eta_{\rm s} \simeq 0.720$, in agreement with earlier analytical thermodynamic predictions~\citep{Kosterlitz1973OrderingMetastability,Halperin1978TheoryTwoDimensionalMelting,Young1979Melting}.  

\begin{figure}
\includegraphics[width=0.5\linewidth]{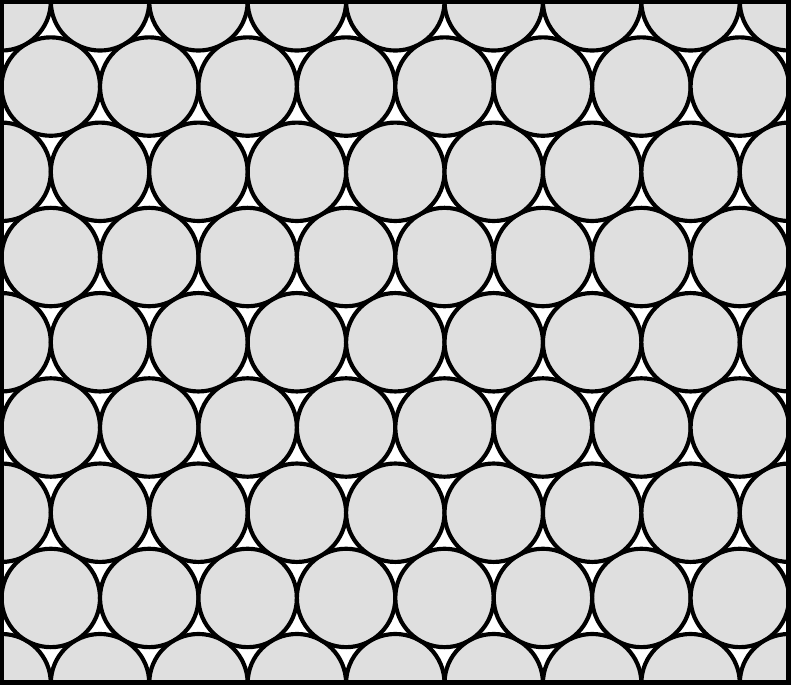}
\caption{A close-packed configuration of hard disks on a two-dimensional torus of volume $(16\sigma) \times (8\sqrt{3}\sigma)$.  This configuration has six-point rotational symmetry about the centre of any disk in the analogous close-packed limit on $\mathbb{R}^2$.}
\label{fig:HardDisksHighDensity}
\end{figure}

\subsubsection{Orientational correlations and non-solid phases.}

The exponentially decaying positional correlations at $\eta = 0.718$ indicate a non-solid phase, but this is not enough to characterise the fluid phase, as the model also possesses \emph{orientational correlations}.  To quantify this, we define the \emph{local orientation} $\Psi_i$ of particle $i$ via the complex number 
\begin{align}
\Psi_i(x) := \frac{1}{| \tilde{S}_i |} \sum_{j \in \tilde{S}_i} \expb{6 i \phi_{ij}}. 
\end{align}
Here $\tilde{S}_i$ is the set of \emph{neighbours} of particle $i$, with particles $i$ and $j$ defined as neighbours if the centre of their minimal separation vector $x_{ij}$ is closer to particles $i$ and $j$ than to any other particle.  The angle $\phi_{ij} \in [0,2\pi)$ is found by expressing $x_{ij}$ in polar coordinates $x_{ij} := (r_{ij},\phi_{ij})$.  The factor of $6$ ensures that the local orientation $\Psi_i(x)$ preserves the six-point rotational symmetry of the close-packed limit (see Figure~\ref{fig:HardDisksHighDensity}).  For a depiction of the local orientation $\Psi_i(x)$ within an example hard-disk configuration, see Figures 1(b-d) of~\cite{Bernard2011TwoStepMelting}.

The \emph{orientational correlation function} is then defined as
\begin{equation}
g_{\rm o}(x; r, \epsilon, \eta, N) := \frac{1}{\mathbb{E} \left| \Psi_i\right|^2}  \sum_{i < j} \mathbb{I} \glc \left| r - \metric(x_i, x_j) \right| < \epsilon \grc \Psi_i^*(x) \Psi_j(x) 
\label{eq:OrientationalCorrelations}
\end{equation}
for all $r > 2\sigma$.  Figure S9 of \cite{Bernard2011TwoStepMelting} and Figure 4.13 of \cite{Bernard2002Thesis} show that for $N = 1024^2$ particles
\begin{align}
\check{g}_{\rm o}(r, \epsilon, \eta, N) \propto 
\begin{cases}
\exp \glb - r / \xi_{\rm o}(\eta) \grb ,  & \textrm{for} \,\, \eta = 0.700, r > 200 \sigma \\
r^{- \alpha_{\rm o}(\eta)} , & \textrm{for} \,\, \eta = 0.718, r > 100 \sigma  \\
\tilde{C} & \textrm{for} \,\, \eta = 0.720, r > 100 \sigma,
\end{cases}
\end{align}
where $\xi_{\rm o}(\eta)>0$ is the \emph{orientational correlation length} of the fluid phase, $\alpha_{\rm o}(\eta) > 0$ is an orientational exponent and $\tilde{C} > 0$ is some constant.  This indicates exponentially decaying orientational correlations for all $r > 200\sigma$ at $\eta = 0.700$, power-law decaying orientational correlations for all $r > 100\sigma$ at $\eta = 0.718$ and non-decaying orientational correlations for all $r > 100 \sigma$ at $\eta = 0.720$. 
These results are consistent with i) an orientationally (and positionally) disordered fluid phase for all $\eta < \eta_{\rm f} \simeq 0.700$, ii) quasi-long range orientational order (and positional disorder) in an \emph{hexatic phase} at $\eta = 0.718$, and iii) long-range orientational order (and quasi-long range positional order) in the solid phase ($\eta > \eta_{\rm s}$). 

\begin{figure}
\includegraphics[width=0.5\linewidth]{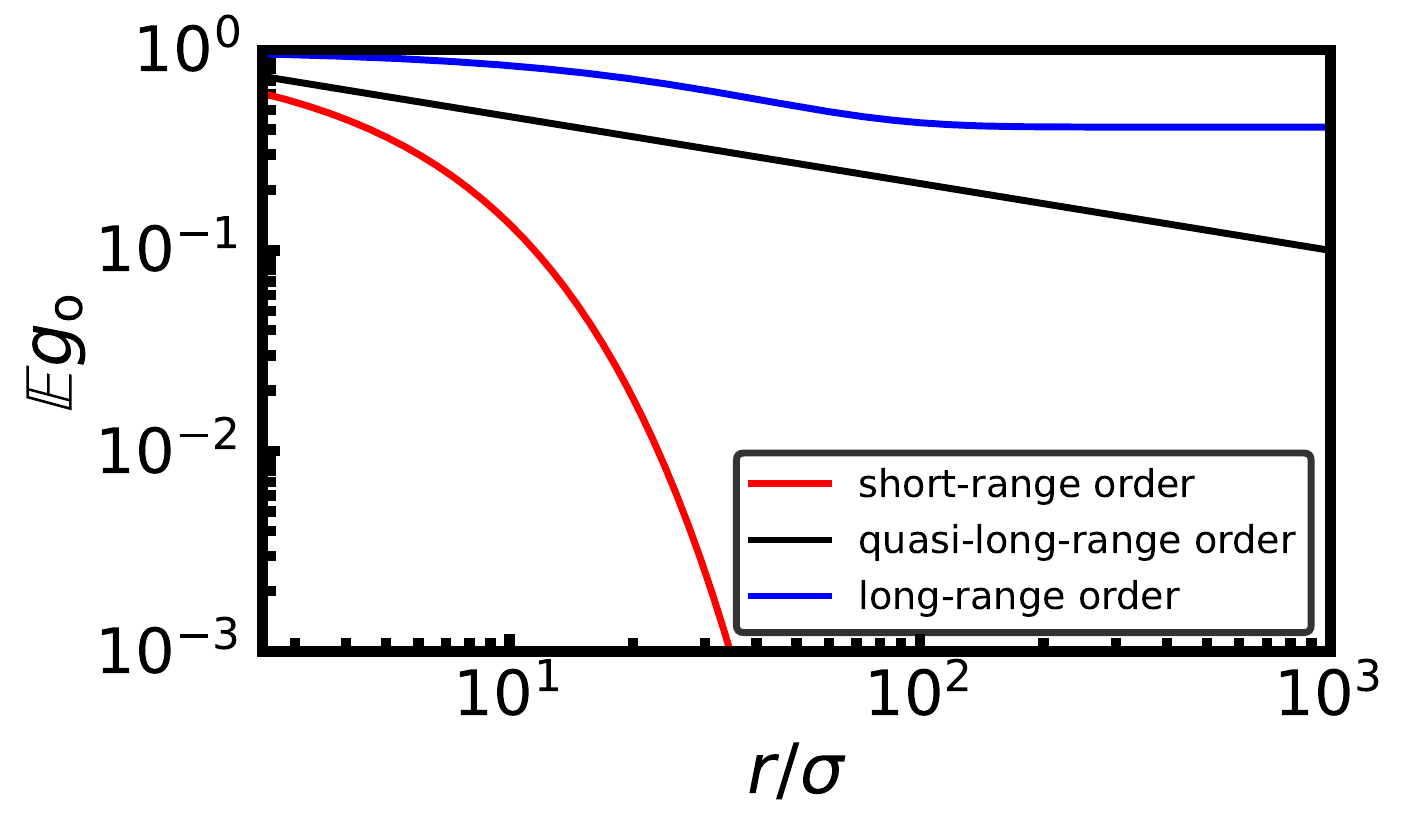}
\caption{A schematic illustration of the three possible forms of long-distance behaviour of the expected orientational correlation function $\check{g}_{\rm o}(r, \epsilon, \eta, N)$ defined in \eqref{eq:OrientationalCorrelations}.  Short-range / quasi-long-range / long-range order corresponds to exponentially decaying / power-law / constant correlations on distances comparable to the linear system size $L$ ($L > 10^3 \sigma$ here).}
\label{fig:CorrelationFunctionSchematic}
\end{figure}

For clarity, Figure~\ref{fig:CorrelationFunctionSchematic} provides a schematic illustration of the three possible forms of long-distance behaviour (short-range order, quasi-long-range order and long-range order) of the expected orientational correlation function.  It is worth noting that the expected positional correlation function also exhibits long-range order (constant positive long-distance correlations) in the close-packed limit, where the particles form the precise hexagonal lattice with six-point rotational symmetry shown in Figure~\ref{fig:HardDisksHighDensity}.  Moreover, in the two-dimensional zero-field Ising model presented in Section~\ref{sec:2DIsing}, an analogously defined spin--spin correlation function exhibits long/short-range order at low/high temperature, and quasi-long-range order at the phase transition.

\subsubsection{Equation of state and fluid--hexatic phase transition.}
\label{sec:CoexistenceInterval}

The above characterises the fluid, hexatic and solid phases, but does not identify a fluid--hexatic phase transition, as this requires analysis of the equation of state \eqref{eq:EqnOfState}.  For the two-dimensional hard-disk model, this is given by~\citep{Metropolis1953EquationOfState,Engel2013HardDiskEquationOfState}
\begin{align}
    \beta \check{p}(\eta, N) = \frac{\eta}{\pi\sigma^2} \glb 1 + 2 \eta \lim_{r \downarrow 2\sigma} \check{g}_{\rm p}(r, \eta, N) \grb .
\end{align}
Equation-of-state calculations then show that, upon transforming to a model in which the pressure $p$ is a hyperparameter, continuously increasing the pressure leads to a discontinuous jump in the expected particle density from $\eta_{\rm f}$ to $\eta_{\rm hex} \simeq 0.716$ at some critical value of the pressure~\citep{Bernard2011TwoStepMelting}.  This is the fluid--hexatic phase transition and the interval $(\eta_{\rm hex}, \eta_{\rm s})$ is the hexatic phase.

\subsection{Lennard--Jones model}
\label{sec:LennardJones}

\begin{figure}
\includegraphics[width=0.5\linewidth]{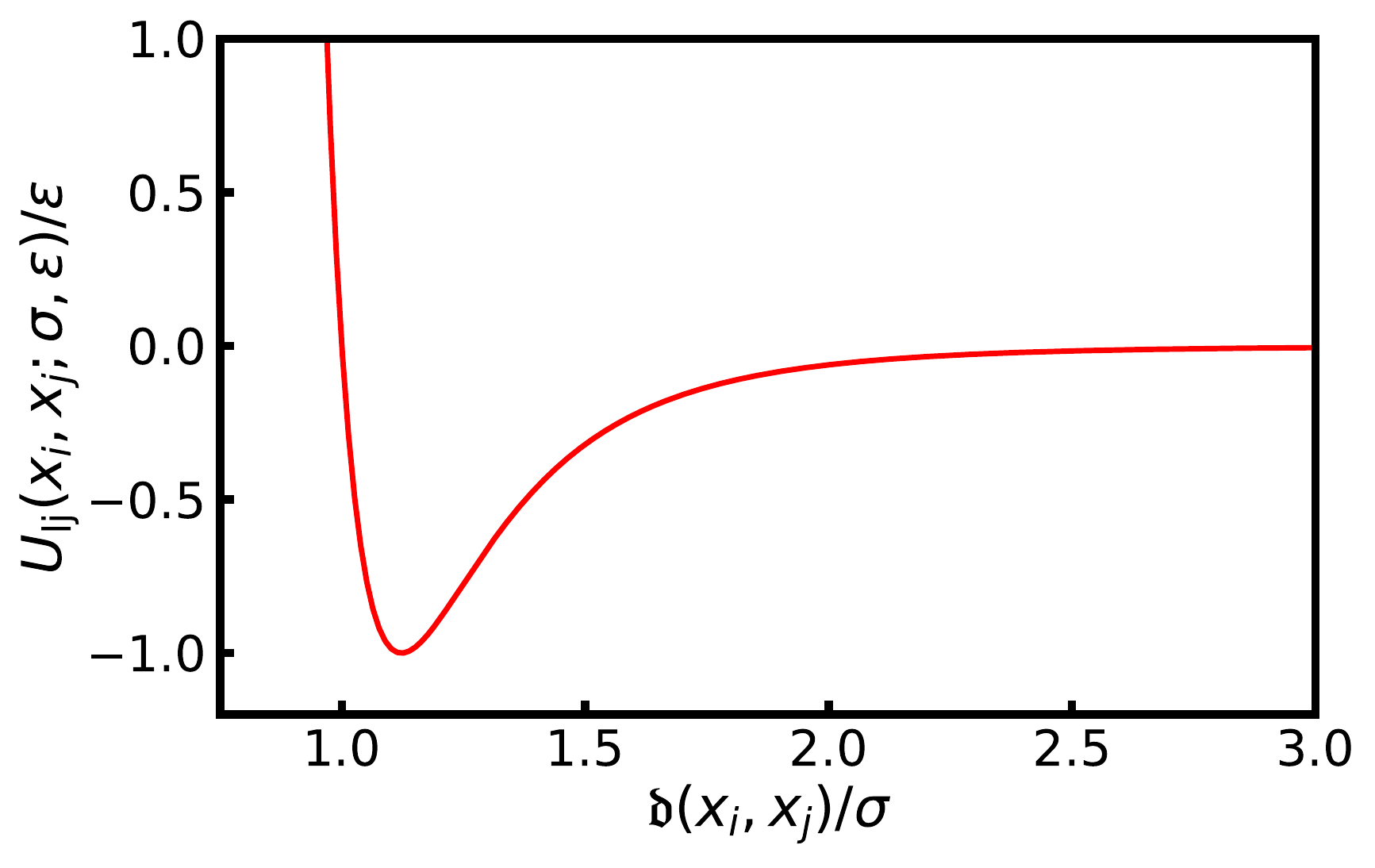}
\caption{Two-particle Lennard--Jones potential.}
\label{fig:LennardJones}
\end{figure}

The Lennard--Jones model describes soft-matter systems composed of $N$ electrically charge-neutral atoms, each on the compact manifold $\mathbb{T}^3$.  It can be viewed as a more sophisticated version of the soft-disk model presented in Section \ref{sec:harddisk} and is defined by the potential $U_{\rm LJ}(x; \eta, \sigma, \varepsilon, N) = \sum_{i < j} U_{\rm lj}(x_i, x_j; \sigma, \varepsilon)$, where 
\[
U_{\rm lj}(x_i, x_j; \sigma, \varepsilon) := 4\varepsilon \left[ \left( \frac{\sigma}{\metric(x_i,x_j)}\right)^{12} - \left( \frac{\sigma}{\metric(x_i, x_j)} \right)^6 \right] 
\]
is the two-particle Lennard--Jones potential between atoms (or particles) $i$ and $j$ (see Figure \ref{fig:LennardJones}).  Here, $\sigma > 0$ determines the most probable minimal separation distance between any two atoms and $\varepsilon > 0$ is the \emph{potential-well depth}.~\footnote{System sizes are typically chosen such that the probability density is negligible where $\metric(x_i, x_j) \sim L$.  This ensures that the periodic boundaries do not qualitatively affect the physics.  For computational efficiency, the two-particle potential is then often set to zero for all $\metric(x_i, x_j) > 2\sigma$.}  The $\metric(x_i,x_j)^{-12}$ \emph{Pauli-repulsion term} is the three-dimensional analogue of a two-particle soft-disk potential and represents the Pauli repulsion between the composite electrons of each atom.  This quantum effect is significant for nearby particles, but diminishes rapidly at larger $\metric(x_i, x_j)$.  The attractive $\metric(x_i,x_j)^{-6}$ \emph{dispersion term} describes the electrical atom--atom attraction due to the instantaneous \textit{electric-dipole moment} of each atom, where the electric-dipole moments are caused by electron-density fluctuations within each atom.  The resultant regions of high electron density within one atom are attracted to resultant regions of low electron density in another.

Using techniques similar to those presented in Section \ref{sec:harddisk}, the model can be studied on its own in order to analyse the liquid--gas phase transition in simple three-dimensional fluids and other physical phenomena.  The two-particle potential is also used as a sub-potential in all-atom models of more complex fluids such as water, as described in Section \ref{sec:allatom}.

\subsection{Electrostatic Coulomb potential}
\label{sec:coulomb}

The toroidal Coulomb potential models electrostatic interactions between $N$ electrically charged particles, each on the compact manifold $\mathbb{T}^3$.  It is derived from the Coulomb law of electrostatics, which states that each Cartesian component of the electrostatic force between particles $i$ and $j$ is proportional to $c_i c_j / \metric(x_i,x_j)^2$, where $c_i \in \mathbb{R}$ is the electric charge of particle $i$~\citep{Coulomb1785SecondMemoire}.  On $\mathbb{R}^3$, the two-particle Coulomb potential is given by 
$$
U_{\rm c} (x_i, x_j; c_i, c_j) = \frac{1}{4\pi \epsilon_0}\frac{c_i c_j}{\metric(x_i, x_j)} ,
$$
with $\epsilon_0 > 0$ the \emph{vacuum permittivity}.  This is the solution of the Poisson equation on $\mathbb{R}^3$.  In simulation, one must use the more involved solution $U_{\rm c} (x_i, x_j; c_i, c_j) \propto c_i c_j G(x_i, x_j)$ to the toroidal Poisson equation $\int_{\mathbb{T}^3} \nabla_x^2 G(x, x') f(x') dx' = - f(x)$ for all test functions $f : \mathbb{R}^3 \to \mathbb{R}$, as presented in \cite{deLeeuw1980SimulationOfElectrostaticSystems}.  Charge neutrality is also required on the torus, either via $\sum_i c_i = 0$ or some charge-neutralisation technique.

If particles $i$ and $j$ have charge values of opposite sign, then $c_i c_j < 0$ and $U_{\rm c}(x_i, x_j; c_i, c_j) \to - \infty$ as $\metric(x_i, x_j) \to 0$, which strongly encourages particles of opposite charge to exist arbitrarily closely together.  When combined with the two-particle Lennard--Jones potential (or a suitable alternative), however, the total potential $U_{\rm c}(x_i, x_j; c_i, c_j) + U_{\rm lj}(x_i, x_j; \sigma, \varepsilon) \to \infty$ as $\metric(x_i,x_j) \to 0$, due to the Pauli-repulsion term.  This combination regularises the Coulomb potential and allows for the simulation of collections of particles with charge values of opposite sign, such as the all-atom model of water presented in Section \ref{sec:allatom}.  We add that, when applied to collections of particles with charge values of the same sign, the electrostatic Coulomb potential can also be studied on its own.

\subsection{Bonded potentials}
\label{sec:bonded}

\begin{figure}[t]
\includegraphics[width=0.49\linewidth]{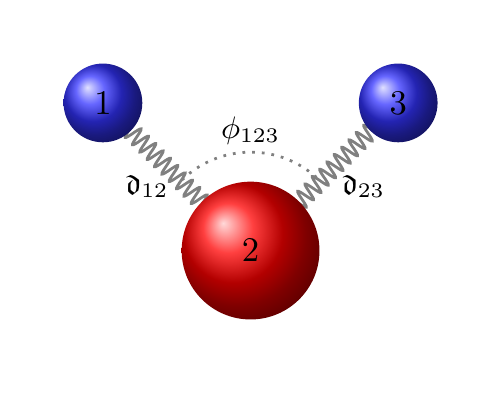}
\caption{Water molecule with indexing as described in Section \ref{sec:allatom}.  Oxygen/hydrogen atoms are red/blue.  The bond angle $\phi_{123} := \phi(x_1, x_2, x_3)$ and the minimal separation distance $\metric_{ij} := \metric(x_i, x_j)$.}
\label{fig:WaterMolecule}
\end{figure}

A molecule is an electrically charge-neutral group of atoms held together by chemical bonds.  Molecular fluids are composed of collections of molecules that interact via intermolecular potentials, such as the Lennard--Jones and Coulomb potentials.  In addition, intramolecular or bonded potentials describe the chemical-bond interactions between the composite atoms of each molecule.  Two of the most common types of bonded potential are bond-stretching and bond-angle potentials, where the former dictate atom--atom minimal separation distances and the latter dictate the angle formed by the positions of three atoms (see Figure \ref{fig:WaterMolecule}).  For three bonded atoms $i$, $j$ and $k$ on $\mathbb{T}^3$, the \emph{harmonic bond-stretching potential} is
$$
U_{\rm s}(x_i, x_j; r_0, k_{\rm b}) := \frac{1}{2} k_{\rm b} \left( \metric(x_i, x_j) - r_0 \right)^2, 
$$
and the \emph{harmonic bond-angle potential} is
$$
U_{\rm a}(x_i, x_j, x_k; \phi_0, k_{\rm a}) := \frac{1}{2} k_{\rm a} \left( \phi(x_i,x_j,x_k) - \phi_0 \right)^2.
$$
Here, $r_0 > 0$, $k_{\rm b} > 0$, $\phi_0 > 0$ and $k_{\rm a} > 0$ are constants that depend on the molecular fluid, and 
$$
\phi(x_i, x_j, x_k) := \arccos \left( \frac{x_{ij}^T x_{jk}}{ \metric(x_i,x_j) \metric(x_j,x_k) } \right) 
$$
is the \emph{bond angle} between three bonded atoms $i$, $j$ and $k$. 

The quadratic bond-stretching potential is derived from \emph{Hooke's law}, but non-quadratic bond-stretching potentials are also used.  For example, graphene typically uses quartic bond-stretching potentials to reflect the enhanced strength of its atomic bonds \citep{wei2011simple}.  Similarly, non-quadratic bond-angle potentials are also used.

\subsection{An all-atom model of water}
\label{sec:allatom}

The electrostatic Coulomb and Lennard--Jones potentials can be combined with the bonded potentials described above to produce an all-atom model of water.  In molecular modelling, an all-atom model is any microscopic model that accounts for the interactions between the individual atoms that form the molecules.  This projects the fundamental quantum-mechanical many-body system onto a simplified classical model of the atomic positions.  Given this significant simplification, many different all-atom models of water exist, where the most suitable model is situation-dependent.  

The simple point-charge water model with flexible molecules~\citep{wu2006flexible} is composed of sub-potentials that describe i) two-body oxygen--hydrogen bond stretching, ii) three-body hydrogen--oxygen--hydrogen angle bending, iii) oxygen--oxygen Lennard--Jones interactions, and iv) electrostatic Coulomb interactions between all (intermolecular) atoms, so that its two-molecule potential is given by 
\begin{align}
U_{\rm mol-mol}(x_1, \dots , x_6) & = \sum_{j \in \{1, 3\}} U_{\rm s}(x_2, x_j) + \sum_{j \in \{4, 6\}} U_{\rm s}(x_5, x_j) + U_{\rm a}(x_1, x_2, x_3) \nonumber \\
+ & \, U_{\rm a}(x_4, x_5, x_6) + U_{\rm lj} (x_2, x_5) + \sum_{i = 1}^3 \sum_{j = 4}^6 U_{\rm c} (x_i, x_j; c_i, c_j) , \nonumber
\end{align}
where particles 2 and 5 are oxygen atoms, particles 1, 3, 4 and 6 are hydrogen atoms, and this two-molecule potential generalises to an arbitrary number of molecules. 
Similar techniques to those presented above are used to analyse the various thermodynamic phases of the model.

\section{Classical sampling algorithms}
\label{sec:ClassicalSamplingAlgorithms}

\subsection{Metropolis}
\label{sec:ClassicalSamplingAlgorithmsMetropolis}

In the first instance of Markov chain Monte Carlo, \cite{Metropolis1953EquationOfState} developed an algorithm to sample from the Boltzmann--Gibbs distribution and applied it to the two-dimensional hard-disk model.  
The algorithm is typically referred to as either `the Metropolis algorithm' or `Monte Carlo' within the statistical physics community. 
Modern-day statisticians, however, may feel more comfortable with the term Metropolis--within--Gibbs, as only a subset of the state is updated at each iteration. Evolving a single particle at each iteration of the algorithm is generally preferred in high-density particle systems, as evolving all particles at once very often leads to slow mixing.


When applied to some $d$-dimensional soft-matter model, each iteration of the Metropolis algorithm consists of proposing the movement of some particle $i$ to a new candidate position $x'_i = x_i \oplus u$, where $x_i \in \mathbb{T}^d$ is the original position of particle $i$, each $u_j \sim \mathcal{U} [-\epsilon, \epsilon]$ for $j \in \{1, \dots, d\}$ for some appropriately chosen $\epsilon > 0$, and $\oplus$ is addition on $\mathbb{T}^d$.  This candidate move is then accepted with probability
$$
\alpha(x_i, x'_i) := \min\left( 1, e^{ -\beta\Delta U(x_i,x'_i) } \right),
$$
where $\Delta U(x_i, x'_i)$ denotes the change in the potential when replacing $x_i$ with $x'_i$.  
This is simply a ratio of Boltzmann--Gibbs distributions. The original algorithm is a systematic scan sampler, meaning the particles are cycled through in a deterministic fashion, rather than randomly selected at each iteration.  When applied to the hard-disk model this probability becomes
\[
\alpha(x_i,x'_i) = \min \left( 1, \frac{\prod_{j \neq i} \mathbb{I} \glc \mathfrak{d}(x'_i,x_j) > 2\sigma \grc}{\prod_{j \neq i} \mathbb{I} \glc \mathfrak{d}(x_i,x_j) > 2\sigma  \grc} \right).
\]
Since this target distribution is uniform, $\alpha(x_i,x'_i)$ simplifies to being one / zero if particles do not / do overlap in the proposed configuration (see Figure~\ref{fig:HardDisks} for examples of accepted and rejected configurations).

Metropolis {\it et al.} investigated the melting transition in two spatial dimensions by simulating the hard-disk model at various choices of disk density.  In the simulations the linear torus size $L = 1$ and the number of particles $N = 224$, and the particle density $\eta$ was varied by adjusting the disk diameter $\sigma$.  At each chosen disk density the initial state was set to be a $14 \! \times \! 16$ hexagonal grid, and the simulations consisted of $16$ burn-in sweeps followed by another $48-64$ sampling sweeps, where a single sweep is $N$ iterations of the algorithm.  Each sweep took around 3 minutes, meaning a total running time of 4-5 hours using the MANIAC computer at Los Alamos National Laboratory.  In fact no evidence of a thermodynamic phase transition was found in the simulations. This is due to the Metropolis algorithm exhibiting extremely slow mixing in the vicinity of both the liquid-hexatic and hexatic-solid phase transitions, because particle moves will very often result in disk overlaps at high particle density, leading to rejections.  To alleviate this, one must choose a very small step size $\epsilon$, which typically leads to very high auto-correlation within the chain and slow convergence to equilibrium.  Similar results were found when applying the Metropolis algorithm to the two-dimensional Lennard--Jones potential and the three-dimensional hard-spheres model in \cite{rosenbluth1954further}.  \cite{wood1957monte} found, however, some evidence of a phase transition when applying the Metropolis algorithm to the three-dimensional Lennard--Jones potential.

\subsection{Glauber dynamics}
\label{sec:ClassicalSamplingAlgorithmsGD}

The Metropolis algorithm can also be applied to the Ising model. At each iteration a candidate move is generated by randomly selecting a particle (meaning a site on the lattice) and flipping the sign of the spin of that particle. This proposal distribution is symmetric and hence the Metropolis rule can be used to accept or reject the move.  Another very similar algorithm introduced in \cite{glauber1963time} and now known as \emph{Glauber dynamics} is also commonly used for this application. 

Glauber dynamics is most easily understood by the statistician as a random scan Gibbs sampler for the Ising model. At each iteration of the algorithm a particle is selected uniformly and a new value for the spin at that site is drawn from the conditional distribution given the spin values of neighbouring particles.

It is natural to compare the two approaches, and in fact this can be done straightforwardly using some well-known tools of the statistician. To do this consider the Glauber dynamics transition as proceeding in three stages. In the first a particle $i$ is randomly selected. In the second a candidate move is considered in which the spin of that particle is changed. In the third the candidate move is accepted with probability
$$
\alpha_{GD}(x_i,x_i') = \frac{e^{-\beta\Delta U(x_i,x'_i)}}{1+e^{-\beta\Delta U(x_i,x'_i)}} .
$$
From this representation, it can be seen that Glauber dynamics can also be viewed as a version of Metropolis--Hastings, whose acceptance rate has been replaced with that advocated by \cite{barker1965monte}.  Using this observation the superiority of Metropolis in terms of asymptotic variance can be established.  The proposition below is an immediate consequence of Theorem 4 in \cite{latuszynski2013clts}.

\begin{proposition}
Let $P_M$ denote the Markov chain produced by the Metropolis algorithm and $P_G$ denote that produced by Glauber dynamics for the Ising model. For any $f$ such that $\sum_{x \in \mathcal{M}^N} f(x)^2 e^{-\beta U_{\textup{Ising}}(x)} < \infty$ it holds that
$$
\nu(P_M, f) \leq \nu(P_G, f) \leq 2\nu(P_M,f) + \textup{Var}_\pi(f)
$$
where $\nu(P, f) := \lim_{n\to\infty} n\textup{Var}(\hat{f}_n)$ is the asymptotic variance of the ergodic average $\hat{f}_n := n^{-1}\sum_{i=1}^n f(X_i)$, with $X_i|X_{i-1}\sim P(X_{i-1},\cdot)$ and $X_1$ a sample from the stationary distribution of $P$.
\end{proposition}


\subsection{Molecular dynamics}
\label{sec:ClassicalSamplingAlgorithmsMD}

In a molecular dynamics simulation Newton's equations of motion are (approximately) solved to directly compute all particle trajectories, after setting random initial velocities.  This approach differs in many ways from the Metropolis algorithm, in which only a single particle is perturbed at each iteration and only the equilibrium behaviour of the system is modelled.  The approach constitutes a direct numerical solution to the $N$-body problem, in order to understand dynamical properties of the system.  The samples produced from certain molecular dynamics simulations can still be used to estimate expected observables at equilibrium, provided the simulation is run for a sufficiently long time.

The molecular dynamics algorithm (abbreviated MD) was first applied to the two-dimensional hard-disk model~\citep{Alder1957PhaseTransition}
before a general method was developed by \cite{Alder1959StudiesInMolecularDynamicsI,Alder1959StudiesInMolecularDynamicsII}.  Unlike the Metropolis algorithm, MD did eventually find convincing evidence of a phase transition in the two-dimensional hard-disk model~\citep{Alder1962PhaseTransition}.  This empirical finding motivated \cite{Kosterlitz1973OrderingMetastability,Halperin1978TheoryTwoDimensionalMelting,Young1979Melting} to develop a two-step theory for the melting transition in two spatial dimensions, which predicted two phase transitions through an intermediate hexatic phase.  Strong particle--particle positional correlations in the vicinity of the transition, however, meant that contemporary molecular dynamics simulations could neither disprove nor corroborate this more nuanced theory. 

\subsubsection{Molecular dynamics for hard disks.}

\begin{figure}
\includegraphics[width=0.5\linewidth]{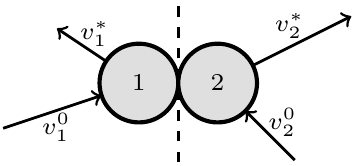}
\caption{A collision of two equal-mass hard disks in a molecular-dynamics simulation.  The dashed line is the perpendicular bisector of the minimal separation vector $x_{21}(t^*) = (2\sigma, 0)$ at the collision time $t^*$ ($x_{21}(t^*)$ is not shown).  The black arrows represent the particle velocities before and after the collision, with arrow length proportional to the norm of the vector.  All changes in velocity are proportional to the component of $v_{21}(t^*) = v_2(t^*) - v_1(t^*)$ perpendicular to the dashed line: $v_1^0 := v_1(t_0) = (1, 1/3) \iota$, $v_1^* := v_1(t^*) = (-1/2, 1/3) \iota$, $v_2^0 := v_2(t_0) = (-1/2, 1/2) \iota$ and $v_2^* := v_2(t^*) = (1, 1/2) \iota$, where $\iota > 0$ has units of velocity.}
\label{fig:HardDisksCollisionMD}
\end{figure}

The hard-disk model is composed of $N$ particles of equal mass, which we set to unity for simplicity.  Newtonian dynamics take a simple form: particles move at constant velocity until two collide, at which point the velocities are updated based on the speed and angle of the collision.  In a typical molecular dynamics simulation the particles are initialised at some chosen positions (e.g. a hexagonal lattice) and given random initial velocities $v_i \in \R^2$ for $i \in \{1,...,N\}$ at time $t_0$ (often the magnitudes are set to be equal and the directions are sampled uniformly).  After this point the system evolves deterministically, with the position of particle $i$ at time $t \geq t_0$ calculated as
\begin{equation}
\label{eq:newton_harddisk}
x_i(t) = x_i(t_0) + (t - t_0) v_i(t), ~~ v_i(t) = v_i(t_0).
\end{equation}
The above equation is correct assuming that no boundary has been crossed. To incorporate periodic boundary conditions the positions must be adjusted modulo $L$ upon hitting a boundary.  Each particle evolves simultaneously in this manner until two collide. The time of the first pair-wise collision between any two particles is again completely pre-determined, and is the first time $t^*$ at which
\begin{equation}
\label{eq:pair_collision}
\mathfrak{d}(x_i(t^*),x_j(t^*)) =  2\sigma
\end{equation}
for some $(i,j)$ pair.  This can be calculated exactly.  Assuming no boundary has been crossed and no other collisions have occurred, the minimal separation vector between particles $i$ and $j$ (i.e. the shortest vector from $x_j$ to $x_i$) is $x_{ij}(t) = x_{ij}(t_0) + (t-t_0)(v_i(t_0) - v_j(t_0))$ for any $t \leq t^*$.  In this case \eqref{eq:pair_collision} becomes a simple quadratic in $t^*$, and denoting $v_{ij}(t) := v_i(t) - v_j(t)$ it has minimum (positive) solution
$$
t^* = t_0 - \frac{x_{ij}(t_0)^T v_{ij}(t_0) + \sqrt{(x_{ij}(t_0)^T v_{ij}(t_0))^2 - v_{ij}(t_0)^2(x_{ij}(t_0) - 4\sigma^2)} }{\| v_{ij}(t_0)\|_2^2},
$$
provided that the expression inside the square root is positive and $x_{ij}(0)^T v_{ij}(0) < 0$.  To account for periodic boundary conditions, the above procedure can be straightforwardly modified by re-evaluating $t^*$ each time a particle passes through a boundary.  
When a collision occurs the velocities of the two involved particles are updated such that the total energy (or the quadratic kinetic energy in this case) is conserved, using the formulae 
\begin{align}
\label{eq:velocity_update1}
v_i(t^*) &= v_i(t_0) - \left(\frac{v_{ij}(t_0)^Tx_{ij}(t^*)}{4\sigma^2}\right)x_{ij}(t^*), \\
\label{eq:velocity_update2}
v_j(t^*) &= v_j(t_0) + \left(\frac{v_{ij}(t_0)^Tx_{ij}(t^*)}{4\sigma^2}\right)x_{ij}(t^*).
\end{align}
Note that $\| x_{ij}(t^*) \| = 2\sigma$ at all collision times $t^*$.  The second term in each equation therefore contains two $x_{ij}(t^*) / \| x_{ij}(t^*) \|$ terms and all changes in velocity are proportional to the component of $v_{ij}$ parallel to $x_{ij}(t^*)$, or perpendicular to the dashed line in the example collision depicted in Figure~\ref{fig:HardDisksCollisionMD}.  After any such collision, $t_0$ is updated to $t^*$ and the process is repeated. 
An example implementation of molecular dynamics for the hard-disk model is given in Algorithm \ref{alg:hard-disk} below.

\begin{algorithm}[H] \label{alg:hard-disk}
\SetAlgoLined
Require $\{(x_i(0),v_i(0)): 1\leq i \leq N \}$, desired collisions $C^* < \infty$\;
Set $t_0 \leftarrow 0$, $C \leftarrow 0$\;
\For{$(i,j) \in \{1,...,N\}^2$ \label{algline:start_iteration}}{
Compute next collision time $t_{ij}^*$\;
}
Set $t^* \leftarrow \min_{(i,j)} t^*_{ij}$\;
\If{$t^* = t^*_{ij}$}{
Update velocities for particles $i$ and $j$ using \eqref{eq:velocity_update1}-\eqref{eq:velocity_update2}\;
}
Set $t_0 \leftarrow t_0 + t^*$, $C \leftarrow C + 1$\;
\For{$i \in \{1,...,N\}$}{
Compute $x_i(t^*)$ using \eqref{eq:newton_harddisk} with boundary corrections\;
}
\If{$C < C^*$}{Return to line \ref{algline:start_iteration};}
 \caption{Molecular simulation for the hard-disk model}
\end{algorithm}

Algorithm \ref{alg:hard-disk} is an exact description of the dynamics of the hark-disk system.  The only numerical errors introduced into molecular dynamics simulation of the model are from floating point arithmetic calculations at collision times.  It may be surprising, therefore, to learn that such errors can sometimes accumulate rapidly.  The reason, put simply, is due to the nature of pair-wise collisions between particles, in which small differences in calculation of the angle of refraction can result in amplified differences in the positions of particles at the next collision.  Further discussion of this phenomenon is provided in Section 2.1.2 of \cite{krauth2006statistical}.  The hard-disk dynamics are often called \emph{event driven}, as ballistic movement of particles is interrupted by collision events. The idea of using event driven dynamics has more recently been applied in statistics to construct sampling algorithms based on Hamiltonian dynamics in the presence of general discontinuous distributions in \cite{nishimura2020discontinuous}.

It is natural to consider the ergodic properties of this approach, which in its simplest form is completely deterministic apart from the random choice of initial particle velocities.  Ergodic properties of various forms of the hard-disk model have now been established under mild conditions by \cite{simanyi2003proof} after pioneering earlier work by \cite{sinai1970dynamical} establishing ergodicity for the case of two particles. The result is significant in forging a concrete connection between Newtonian dynamics (on which the equations of motion are based) and the Boltzmann distribution (from which the equilibrium properties of the system are deduced). Among mathematicians, models such as the hard-disk system are often referred to as the study of dynamical billiards (e.g. \cite{tabachnikov2005geometry}).

\subsubsection{Smooth potentials: the microcanonical ensemble.}

Newton's equations of motion for smooth potentials are given by the dynamical system
\begin{equation}
\label{eq:newton}
m_i\frac{d^2x_i}{dt^2} = F_i(x),
\end{equation}
where $m_i \in [0,\infty)$ denotes the mass of particle $i$ and $F_i(x) := -\nabla_i U(x)$ is the total force acting on particle $i$.  Here 
$\nabla_i$ is the gradient operator of particle $i$.  It is common to introduce an auxiliary velocity variable $v_i := \dot{x}_i$, reducing the second order system \eqref{eq:newton} into a first order system
\begin{equation}
\label{eq:hamilton_dynamics}
\begin{aligned}
\dot{x}_i &= v_i, \\
\dot{v}_i &= F_i(x)/m_i.
\end{aligned}
\end{equation}
The above dynamics can also be described in terms of position and momenta $p_i := m_i v_i$.  This \emph{Hamiltonian} formulation of classical mechanics gives rise to a dynamical system with several appealing features, such as volume preservation and invariance of the Hamiltonian function $H(x, p) = U(x) + \sum_i p_i^T p_i/ (2 m_i)$, which describes the total energy of the system by combining the potential $U(x)$ with a quadratic kinetic energy term.  

The system \eqref{eq:hamilton_dynamics} cannot usually be solved analytically, but in many cases numerical integrators that preserve many geometrical features of the original system exist.  A general survey is beyond the scope of this article, but see \cite{hairer2003geometric,bou2018geometric} for comprehensive reviews or \cite{leimkuhler2004simulating,hairer2006geometric} for book-length treatments.  The most popular algorithm in use today is the \textit{velocity Verlet} algorithm \citep{verlet1967computer}, in which the dynamics are approximated by first taking a half-step in the momentum component $p_i(\varepsilon/2) = p_i(0) + (\varepsilon / 2) F_i(x(0))$ for each particle (where $\varepsilon>0$ is the step-size) and then iterating the \emph{leapfrog} dynamics for each $n \in \{1,...,\ell\}$ 
\begin{equation} \label{eq:leapfrog}
\begin{aligned}
x_i(n\varepsilon) &= x_i((n - 1)\varepsilon) + \varepsilon p_i((n - 1/2)\varepsilon)/m_i \\
p_i((n + 1/2)\varepsilon) &= p_i((n-1/2)\varepsilon) + \varepsilon F_i(x(n\varepsilon)),
\end{aligned}
\end{equation}
before a half-step in the momentum component is taken at the final iteration, in order to generate a skeleton trajectory up to time $\ell\epsilon$ (the alternative \textit{position Verlet} algorithm begins and ends with a half-step in the position). The momentum update is referred to as the `kick' and the position update the `drift'.  The algorithm is also known as the \emph{leapfrog} scheme, owing to the intermediate leapfrogging action of the position and momentum coordinates, and is popular because only one force evaluation $F(x)$ is needed per time step (ignoring the initial and final kicks) while achieving $\mathcal{O}(\varepsilon^2)$ global error over fixed time scales \citep{bou2018geometric}. The algorithm was originally used within the contemporary molecular dynamics literature by Loup Verlet to simulate a system of 864 particles interacting under a Lennard--Jones potential, but had been used much earlier than this by \cite{stormer1907trajectoires}, and is sometimes called St\"{o}rmer--Verlet integration for this reason.  The positions of particles are then updated again after each leapfrog step to incorporate periodic boundary conditions by applying the modular transformation described in Section~\ref{sec:PeriodicBoundaries}.

The molecular dynamics algorithm described above is restricted to exploring the \emph{microcanonical ensemble}, meaning the space of possible states
$$
\{ (x,p) \in \mathcal{M}^N \times \mathbb{R}^{Nd} : H(x,p) = H_0 \},
$$
where $H_0 := H(x(0),p(0))$, combined with the uniform probability measure over these states.  The total energy of the system hence remains constant.  On configuration space $\mathcal{M}^N$, this usually will not correspond to the support of the Boltzmann--Gibbs distribution $e^{-\beta U(x)}$, as the potential is restricted to the set $\{U(x) \leq H_0\}$ owing to the non-negativity of the kinetic energy. This was of no concern for the hard-disk model as the potential is almost everywhere constant, meaning that each level set of $H$ allows exploration of the entire space. For general potentials, this requires the ability to move between contours of the Hamiltonian. The collection of states
$$
\{ (x,p) \in \mathcal{M}^N \times \mathbb{R}^{Nd} : e^{-\beta H(x,p)} >0 \} 
$$
combined with the probability measure $\propto e^{-\beta H(x,p)}dxdp$ over these states, is known as the \emph{canonical ensemble}.  In some physical settings the microcanonical ensemble is of direct interest, but if the canonical ensemble is desired then the above approach to molecular simulation is no longer sufficient.

\subsubsection{Smooth potentials: the canonical ensemble.}

There are various approaches to simulating the canonical ensemble, and hence exploring the entirety of the Boltzmann--Gibbs distribution $\pi(x) \propto e^{-\beta U(x)}$. From the physical perspective sampling from the canonical ensemble can be understood as allowing the system of particles to exchange energy with the outside world.  The system is often assumed to be contained within a \emph{heat bath} or \emph{thermal reservoir} meaning that the system temperature $\beta^{-1}$ can be controlled whilst still allowing for heat exchange with the external environment.  Here we will primarily discuss \emph{Langevin dynamics} to sample from the canonical ensemble, although there are many other approaches for this task \cite[Chapter~6]{leimkuhler2016molecular}.  

Langevin dynamics consist of adding some stochasticity to Newton's equations of motion \eqref{eq:newton}, which allows the total system energy to fluctuate over time. The deterministic system \eqref{eq:hamilton_dynamics} is combined with an Ornstein--Uhlenbeck process on the momentum coordinate, resulting in the system of stochastic differential equations
\begin{equation}
\label{eq:udlangevin}
\begin{aligned}
dx_i(t) &= m_i^{-1}p_i(t)dt \\
dp_i(t) &= F_i(x(t))dt - \gamma m_i^{-1}p_i(t)dt + \sqrt{2\gamma \beta^{-1}}  dW_i(t) ,
\end{aligned}
\end{equation}
where each $(W_i(t))_{t\geq 0}$ is a standard Wiener process on $\mathbb{R}^d$, and $\gamma>0$ controls the strength of frictional forces.  The linear drift term $-\gamma m_i^{-1}p_i(t)dt$ taken in isolation results in an exponential decay in the momentum, with $\gamma$ dictating the rate at which energy dissipates from the system due to friction.  The final term $dW_i(t)$ represents an injection of stochastic force, and its coefficient can be determined using the \emph{fluctuation-dissipation theorem} \cite[Chapter~9]{pavliotis2014stochastic}.  We direct the interested reader to Section 6.3.2 of \cite{leimkuhler2016molecular} for more precise physical intuition.

The system \eqref{eq:udlangevin} is commonly known as \emph{underdamped} Langevin dynamics.  It can be solved numerically in various ways.  One popular approach is to split the dynamics into two separate systems, the first being simply the Hamiltonian system \eqref{eq:hamilton_dynamics}, and the second the Ornstein--Uhlenbeck process $dp_i(t) =  - \gamma m_i^{-1}p_i(t)dt + \sqrt{2\gamma \beta^{-1}}  dW_i(t)$, which has explicit solution
\begin{equation} \label{eqn:outransition}
p_i(t)|p_i(0) \sim N\left( p_i(0) e^{-\gamma m_i^{-1} t}, m_i\beta^{-1} (1-e^{-2\gamma m_i^{-1} t}) \right).
\end{equation}
The full system \eqref{eq:udlangevin} can then be approximated by iterating between numerically solving \eqref{eq:leapfrog} for a short time increment and exactly solving the Ornstein--Uhlenbeck dynamics for the same length of time.  Justification for solving in this manner is given by the Baker--Campbell--Hausdorff formula (e.g. \cite{leimkuhler2004simulating}).  Ergodic properties and mixing times of underdamped Langevin dynamics are an area of constant study in the applied mathematics literature (e.g.~\cite{wu2001large,talay2002stochastic,mattingly2002ergodicity}) as well as the behaviour of discretisation schemes (e.g.~\cite{mattingly2002ergodicity,durmus2021uniform}).  Interest in the approach has also grown recently within the machine learning community in the context of establishing non-asymptotic mixing-time bounds for approximate sampling algorithms (e.g.~\cite{dalalyan2020sampling}).

A simplified form of the above dynamics that will be very familiar to statisticians and machine learners can be found by considering a particular limiting regime of \eqref{eq:udlangevin}.  Taking $m_i = 1$ for simplicity, in the \emph{overdamped limit} $\gamma \to \infty$ the right-hand side of \eqref{eqn:outransition} becomes $N(0, \beta^{-1})$, thereby flushing all memory of previous momenta from the system. Alternating \eqref{eqn:outransition} and \eqref{eq:leapfrog} in this limit results in the transition $x_i(\varepsilon) = x_i(0) + \varepsilon^2F_i(x(0))/2 + \varepsilon \zeta_i$, where $\zeta_i \sim N(0,\beta^{-1})$, which is the Euler--Maruyama numerical scheme applied to the stochastic differential equation
\begin{equation}
\label{eq:odlangevin}
dy_i(t) = F_i(y(t))dt + \sqrt{2\beta^{-1}}dB_i(t),
\end{equation}
where $y(t) := x(\gamma t)$.
A rigorous derivation of the above is provided in Section 6.5 of \cite{pavliotis2014stochastic}, see also Section 1.2.2 of \cite{stoltz2021computational}. Among the statistical physics community \eqref{eq:odlangevin} is often called \emph{Brownian dynamics}, as introduced by \cite{rossky1978brownian}. Among statisticians \eqref{eq:odlangevin} is the starting point of the Metropolis-adjusted Langevin algorithm popularised by \cite{roberts1996exponential} and used extensively since this point.

\subsection{Hybrid algorithms}

We end this section by briefly mentioning hybrid Monte Carlo, which is now more commonly known as Hamiltonian Monte Carlo \citep{neal2011mcmc}.  Following the above discussion the original name should seem natural, as among physicists the algorithm is easily understood as a hybrid of the molecular dynamics and Metropolis approaches.  The algorithm was introduced in lattice field theory by \cite{duane1987hybrid}, and was popularised in the statistics literature by \cite{neal1993bayesian}.  Today it is considered to be among the state-of-the-art approaches to sampling in many statistical applications. Two notable examples where the method has found success are posterior distributions for Bayesian neural networks \citep{neal2012bayesian} and hierarchical models \citep{betancourt2015hamiltonian}.

Today it is widely used in both disciplines and has been extensively studied.  Indeed, many recent algorithmic innovations have been made by statisticians and machine learners, such as the automated tuning of the time for which Hamilton's equations should be simulated before a Metropolis step is applied and the momentum is re-sampled \citep{hoffman2014no,sherlock2021apogee}, and the incorporation of geometric ideas to allow for position-dependent masses that use local information about $\pi$ for sampling highly anisotropic distributions \citep{girolami2011riemann,betancourt2017geometric}.

\section{Advanced algorithms}
\label{sec:AdvancedAlgorithms}

\subsection{Cluster algorithms for lattice models}
\label{sec:ClusterAlgorithms}

The potential associated with the Ising model induces strong correlations between spin values at neighbouring lattice sites when the inverse temperature $\beta$ is large. This can make the sampling task very challenging using a site-by-site updating strategy as employed by the Metropolis algorithm and Glauber dynamics, leading to poorly mixing Markov chains.  An alternative approach is to consider changing the spins of several particles in a single step of the algorithm.  A popular strategy for doing this was proposed by \cite{swendsen1987nonuniversal}, and later modified by \cite{wolff1989collective}.  In the following we set $h = 0$ (see \eqref{eq:1dIsingFreeEnergy}) for brevity.

The Swendsen--Wang algorithm is based on the idea of changing the spin values of entire clusters of particles together.  The key ingredient is to introduce an auxiliary variable for each edge joining adjacent lattice sites (excluding edges passing through the periodic boundaries).  Consider a lattice of $N$ sites (corresponding to $N$ particles) and label each with an index $1 \leq i\leq N$. Each edge in the lattice can then be assigned an auxiliary \emph{bond} variable indexed by the two particles that it connects. In the one-dimensional Ising model this equates to introducing $N-1$ auxiliary variables, and for a $d$-dimensional lattice $dN^{(d-1)/d}(N^{1/d}-1)$ such variables. For two neighbouring particles $i$ and $j$ we denote the associated bond variable $b_{ij} \in \{0,1\}$. If $b_{ij} = 1$ then particles $i$ and $j$ are grouped in the same cluster, and if not they belong to different clusters. The edge variables therefore partition the set of particles. If $x_i \neq x_j$ then $b_{ij} = 0$, meaning only particles with the same spin can be in the same cluster. If $x_i = x_j$ then $b_{ij} = 0$ with probability $e^{-2\beta J}$, meaning
$$
\mathbb{P}[b_{ij} = 1|x_i, x_j] = q_{ij}(x) :=  1- \exp\left\{ -2\beta J \mathbb{I}(x_i = x_j) \right\}.
$$
Once all edge variables have been sampled, all spins within each cluster are flipped with probability $1/2$, and for each cluster the decision of whether or not to flip the spins is taken independently of all other clusters. In this way, large numbers of particles can be flipped simultaneously.  A proof of the following well-known result is provided in the supplement \citep{faulkner2023supplement}.

\begin{proposition} \label{prop:swendsenwang}
The Markov chain induced by the Swendsen--Wang algorithm targeting the Boltzmann--Gibbs distribution $\pi(x) \propto e^{-\beta U_\textup{Ising}(x; J, 0, N)}$ is ergodic for any choice $\beta \in (0,\infty)$, $J > 0$ and $N \in \mathbb{N}$.
\end{proposition}

Conditions for rapid mixing of the algorithm are discussed in \cite{gore1999swendsen}, and convergence has also been considered by \cite{huber2003bounding}. Generalizations and further discussion are provided in \cite{edwards1988generalization} and elsewhere.

The Wolff algorithm \citep{wolff1989collective} differs from the approach of Swendsen \& Wang in that only a single cluster is flipped at each iteration. The approach can be uncovered by sampling each bond variable as in the Swendsen--Wang algorithm, but then simply choosing a particle uniformly at random and flipping the spins within the cluster to which that particle belongs.  There is, however, another mathematically equivalent way to construct the cluster to be flipped that does not require every bond variable to be sampled, and is therefore computationally more efficient.  We do not provide details here but refer the interested reader to Section 5.2.3 of \cite{krauth2006statistical}.  \cite{wolff1989comparison} reports superiority of the single cluster approach through simulations on a $64^3$ lattice near the critical temperature.

\cite{nott2004bayesian} applied the Swendsen--Wang approach to Bayesian variable selection. In Bayesian variable selection auxiliary variables $\gamma_j \in \{0,1\}^p$ are often introduced, allowing a spike-and-slab prior to be placed on each $\beta_j$ by specifying that if $\gamma_j = 0$ then $\beta_j = 0$ with probability 1, and otherwise $\beta_j|(\gamma_j = 1)$ has a continuous prior distribution. When appropriate priors are chosen the marginal posterior distribution for $\gamma \in \{0,1\}^p$ can often be obtained directly, meaning that Markov chain Monte Carlo methods can be employed directly on this space. As in the Ising model, the result is a distribution to be sampled from that is defined over a large space of correlated random variables that can each take two possible values. \cite{nott2004bayesian} found that the Swendsen--Wang approach can yield substantial improvements compared to component-wise Metropolis in the presence of high multi-collinearity.  Improvements were not always observed, however.  The models are not of course identical.  In particular the correlations between $\gamma_j$ variables in Bayesian variable selection are dictated by the data, and therefore some can be high and others low in an unstructured manner. This contrasts with the rigid correlations imposed by the lattice structure of the Ising model discussed here.

Following the success of the cluster approach, similar ideas were used to design algorithms for the hard-disk and other soft-matter models \citep{dress1995cluster}.  A different strategy has, however, proved more successful in these systems, which we turn to next.

\subsection{Jaster's algorithm for hard disks and spheres}

A key weakness of the Metropolis algorithm applied to the hard-disk model at high particle density is that randomly perturbing the position of a particle is very likely to cause overlap with another, leading to a rejected move.  
\cite{jaster1999computer,jaster1999improved} proposed a simple approach to combat this. In Jaster's algorithm an initial uniform innovation $u$ is drawn and a particle $i$ is selected uniformly at random.  The first proposal is then to move particle $i$ to $x_i \oplus u$, as in the Metropolis algorithm.  If this results in an overlap with particle $j$, then a new position $x_j \oplus u$ is proposed for this particle.  The process continues until either a configuration is found in which no particles overlap, a particle overlaps with more than one other, or a pre-specified maximum number of attempted moves have been made without finding a new non-overlapping configuration. If the first of these three scenarios occurs then the new configuration is accepted, and in either of the others it is rejected.

Jaster's algorithm can be described as a particular case of the delayed-rejection algorithm \citep{mira2001metropolis} from statistics.  It can also be cast in the more recent sequential proposal Markov chain Monte Carlo framework of \cite{park2020markov}. Full details of this are given in Section 7.5 of \cite{andrieu2020general}.

Jaster's algorithm is an improvement on Metropolis in the sense that the probability of rejection is strictly lower. This in turn improves the asymptotic variance of the resulting Markov chain per iteration (e.g. \cite{mira2001ordering}), although each iteration is now also more expensive.  The improvements, however, are often small in the case of high-density particle systems, in which it is very likely that the first stage proposal will result in the active particle $i$ overlapping with more than one other. Jaster acknowledges this and suggests a modification in which the same particle is moved by very small amounts in one direction at each iteration, in order to reduce the chances of multiple particle overlaps. The idea has since been fully developed and will be introduced in the next section.

\subsection{Event chain Monte Carlo}
\label{sec:ecmc}

Event chain Monte Carlo can be viewed as a natural innovation of Jaster's algorithm to alleviate the issue of collisions involving more than two disks. The central idea is embedded in Jaster's remark that if disks are perturbed by smaller increments then configurations involving multiple disk overlaps are less likely to occur.  The same logic suggests that the limiting continuous-time algorithm in which a single particle makes infinitesimally small moves (until collision) would completely remove the danger of collisions involving more than two particles. The resulting sampling algorithm applied to hard-disks is the event chain Monte Carlo algorithm of \cite{Bernard2009EventChain}, which was later extended to general potentials by \cite{michel2014generalized}.

The event chain Monte Carlo algorithm simulates a continuous-time stochastic process known as a piecewise deterministic Markov process (PDMP), which involves deterministic dynamics and jumps, but no diffusive behaviour. This form often lends itself well to exact simulation. PDMPs have also been proposed as sampling algorithms in statistics, notably by \cite{BouchardCote2018BouncyParticleSampler,bierkens2019zig}. A recent review is given in \cite{fearnhead2018piecewise}.

\subsubsection{Event chain Monte Carlo for the hard-disk model.}

To simulate event chain Monte Carlo for the hard-disk model an initial configuration and active particle $i \in \{1,...,N\}$ must be chosen. A single-particle velocity $u$ is then simulated from some initial two-dimensional distribution on the unit circle. The configuration of the process $x(t)$ at time $t \geq 0$ is then determined in a manner that has some parallels with molecular dynamics simulations for the hard-disk model. The active particle $i$ moves at unit speed in the direction $u$ while all other particles remain still. The time to the first collision can then be calculated using \eqref{eq:pair_collision} as in a molecular dynamics simulation. When a collision occurs the active particle is updated. 

This can be described mathematically by defining the flow operator
\begin{equation} \label{eq:ecmc_flow}
\phi_t(x,v,i) := (x+tv,v,i)
\end{equation}
for configuration $x$ with $N$-particle velocity vector $v \in\mathbb{R}^{2N}$, which will have only two non-zero entries corresponding to the velocity $u$ of the active particle $i$.  We define the colliding particle as 
$$
c(x,i) := 
\begin{cases} 
\arg\min_{j\neq i} \metric(x_i,x_j)  & \text{if } \min_{j\neq i}\metric(x_i,x_j)\leq 2\sigma \\
i & \text{otherwise}.
\end{cases}
$$
It is of course still possible that two particles could be equidistant from $i$, meaning $c(x,i)$ takes multiple values. At equilibrium this is a measure-zero event, but care must be taken to initialise the algorithm so that this does not occur.  Upon collision the velocity can be updated in numerous ways. Of course it must be transferred such that the active particle is now $c(x,i)$. But the non-zero part of the velocity $u$ can also be modified. In \emph{reflected event chain Monte Carlo} it is adjusted according to the angle of the collision, as in a molecular dynamics simulation. This strategy did not perform as well, however, as the version known as \emph{straight event chain Monte Carlo} in which $u$ is not modified after collisions.  In this case the new $N$-particle velocity $s(v,i,c(x,i))$ is calculated by simply swapping the $i$th and $c(x,i)$th two-dimensional components of $v$, which results in the $c(x,i)$th component becoming $u$ and the $i$th component becoming zero.  The \emph{swap-upon-collision} operator can therefore be defined as
$$
S(x,v,i):= (x, s(v,i,c(x,i)), c(x,i)).
$$
Iterating the maps $\phi_\varepsilon$ for some small $\varepsilon > 0$ and $S$ then leads to an algorithm in which a particle moves in a straight line until collision, at which point the active particle and velocity vector are updated.  Defining the augmented state $z :=(x,v,i)$, event chain dynamics can therefore be formally defined as $z(t) := \xi_t(z(0))$, where 
$$
\xi_t(z(0)) := \lim_{\varepsilon \to 0} [S\circ\phi_\varepsilon]^{\lfloor t/\varepsilon \rfloor}(z(0)) 
$$
for any $t \geq 0$, which is a completely deterministic trajectory through time.

The event chain algorithm also involves a final `refreshment' step, which plays a similar role to momentum refreshment in hybrid Monte Carlo.  At certain times the single-particle velocity $u$ is changed in some way.  This is to help the process reach equilibrium. This refreshment can be done in several ways, one example being complete uniform re-sampling from the unit circle.  In the best-performing implementation, however, the initial value of $u$ is chosen to be either $(1,0)$ or $(0,1)$, each with probability $1/2$, and each refreshment simply entails swapping the elements of $u$ \citep{Bernard2009EventChain}. This implementation is called the $xy$-version of event chain Monte Carlo, as the active particle will always be travelling parallel to either the $x$- or $y$-axis.

If the single-particle velocity refreshments are implemented according to the $xy$-version of the algorithm at fixed times, then the event chain algorithm for hard-disk systems is in fact completely deterministic. If, however, the refreshment times occur according to a Poisson process, then some stochasticity is introduced. 

\subsubsection{(Generalized) Event chain Monte Carlo for smooth potentials.}

In the hard-disk model the form of the potential leads to a natural definition of both collisions and updates of the active particle.  It is not immediately obvious, however, how to extend these ideas to sampling from continuous potentials.  Fortunately these hurdles were overcome by \cite{michel2014generalized} following earlier work from \cite{Peters2012RejectionFreeMonteCarlo}.


We will consider the case of a potential composed of generic pairwise components $U_{\rm g}(x_i,x_j)$.  As in the case of hard-disks let $z:= (x,v,i)$ be the augmented state of the process. The algorithm proceeds by defining a Poisson process for each pair of particles. The process associated with particles $j$ and $k$ has event rate defined by the function
$$
\lambda_{jk}(z) := \beta \mathbb{I}(i = j) \max\left( 0, \langle \nabla_x U_{\rm g}(x_j, x_k), v \rangle \right).
$$
This will only be non-zero if $j$ is the current active particle. 
If an event associated with this process occurs then the active particle is swapped from $j$ to $k$. As in the hard-disk model this is achieved by swapping the $j$th and $k$th $d$-dimensional components of the velocity vector $v$. In the absence of an event the process advances deterministically according to the flow map \eqref{eq:ecmc_flow}, which results in the active particle moving with constant velocity $v$ with all others remaining in place. The algorithm also includes a velocity refreshment step.

\subsubsection{The generator and its properties.} 
\sloppy Recall that the infinitesimal generator of a continuous-time Markov process is defined point-wise as $Lf(z):= \lim_{\delta \to 0} \delta^{-1}\left( \mathbb{E}[f(z(t+\delta))|z(t) = z] - f(z) \right)$, whenever such a limit is well-defined (we will skip the technicalities here in favour of a more intuitive discussion).  The generalized event chain algorithm can be described through the PDMP with infinitesimal generator
\begin{equation} \label{eq:ECgenerator}
Lf(z) := \langle v, \nabla_x f(z) \rangle + \sum_{j < k} \lambda_{jk}(z)[f(S_{jk}(z))-f(z)] + \lambda_{\rm ref}\int [f(z') - f(z)]R(z, dz')
\end{equation}
where
$$
S_{jk}(z) := (x, s(v, j,k), k)
$$
and the function $s(v,j,k)$ swaps the $j$th and $k$th ($d$-dimensional) elements of $v$. In the final term on the right-hand side $\lambda_{\rm ref} >0$ denotes the refreshment rate for the active velocity component, and $R(z, dz')$ is a Markov kernel only changing the active component of the velocity.  In different implementations $R$ can either perform uniform refreshment for $v_i = u$ on the unit sphere or apply the $xy$ transformation, which in the $d$-dimensional case we will treat as shifting each element of $u$ one space to the right modulo $d$.

Properties of the Markov process associated with the event chain algorithm can be extracted by studying \eqref{eq:ECgenerator} as an operator on a suitably defined Hilbert space. Here we will consider the space $L^2(\mu)$, where $\mu(dz)$ is the product measure formed by combining $\pi(dx)$ with the discrete uniform distribution on $\{1,...,N\}$ for $i$ and a uniform distribution on the $(Nd-1)$ unit sphere for $v$ (an alternative choice is the conditional probability measure for $v|i$ for which the $i$th $d$-dimensional component of $v$ is uniform on the unit $(d-1)$-sphere and all other elements are $0$).  Recall that on such a space a $\mu$-reversible process will have the property that the associated generator is self-adjoint, meaning $\langle f, Lg\rangle_\mu = \langle Lf, g\rangle_\mu$, where $\langle f,g\rangle_\mu := \int f(z)g(z)\mu(dz)$ is the $L^2(\mu)$ inner product. The operator \eqref{eq:ECgenerator} is not self-adjoint, meaning the process is not $\mu$-reversible. We can, however, consider a more general property introduced in \cite{andrieu2021peskun} known as $(\mu,Q)$-self-adjointness. This means that there is another operator $Q$ on $L^2(\mu)$ satisfying $Q^2 = I$ and $\langle f, g \rangle_\mu = \langle Qf, Qg\rangle_\mu$ (called an \emph{isometric involution}) and for which
\begin{equation} \label{eq:Qsa}
    \langle f, Lg \rangle_\mu = \langle QLQf, g\rangle_\mu.
\end{equation}
Clearly making the choice $Q = I$ equates \eqref{eq:Qsa} with $\mu$-reversibility, but other choices are possible. It is shown in Appendix E of \cite{andrieu2020general} that the generator associated with event chain Monte Carlo satisfies \eqref{eq:Qsa} with $Qf(x,v,i) := f(x,-v,i)$ when the refresh kernel $R$ is taken to be uniform. We extend this to the $xy$ implementation below when $d=2$ (a proof is provided in the supplement \citep{faulkner2023supplement}).  For this particular choice $(\mu,Q)$-self-adjointness can be related to the notion of skew-detailed balance for a discrete time Markov chain with a velocity component (e.g. \cite{vucelja2016lifting,turitsyn2011irreversible}), since the associated transition kernel $P_t(z,dz)$ satisfies skew-detailed balance for any choice of $t \geq 0$ (see Theorem 9 in \cite{andrieu2021peskun}).

\begin{proposition} \label{prop:ECqsa}
The event chain Monte Carlo infinitesimal generator \eqref{eq:ECgenerator} with $xy$-refreshments is $(\mu,Q)$-self-adjoint with the choice $Qf(x,v,i) := f(x,-v,i)$ when $d=2$.
\end{proposition}

\begin{remark} \label{rem:ergodicityECMC}
Note that Proposition \ref{prop:ECqsa} is not wholly satisfactory because in the $xy$-version of event chain Monte Carlo the active velocity component $u$ only ever takes one of $d$ values on the $(d - 1)$-dimensional unit sphere (in contrast with uniform refreshment on the unit sphere).  The algorithm therefore satisfies $(\mu,Q)$-self-adjointness with the indicated $\mu$ when $d=2$, but is not $\mu$-irreducible, meaning that $\mu$ is not the limiting distribution for the chain. More generally no rigorous proof of ergodicity for the $xy$-version of the straight event chain algorithm is known to the authors at the time of writing.
\end{remark}

\begin{remark}
The potential associated with the hard-disk model is not smooth, meaning a generator-level definition akin to \eqref{eq:ECgenerator} is not straightforward. One route to such an object is to consider a sequence of processes associated with the soft-disk potential \eqref{eq:SoftDiskModel} indexed by $k$, and then letting $k \to \infty$. Similar ideas underpin the extension of Hamiltonian Monte Carlo to non-smooth models discussed in \cite{nishimura2020discontinuous}. Recent work by \cite{monemvassitis2023pdmp} has provided a generator-level description via a different approach and established conditions under which ergodicity can be shown for the uniform refreshment process.
\end{remark}

\subsubsection{Implementation details.}

The generator-level description given in \eqref{eq:ECgenerator} assumes that the potential can be broken into factors at the pair-wise level, but in practice other factorisation schemes are possible. In the case of pair-wise factorisation the continuous-time algorithm can be derived by taking an appropriate limit of a discrete-time Metropolis algorithm in which the usual acceptance rate is replaced with the \emph{factorized Metropolis filter} $\prod_{i\neq j} \min(1, e^{-\beta ( U_{\rm g}(x'_i,x'_j) - U_{\rm g}(x_i,x_j) )})$ \citep{michel2014generalized}. This is of course less efficient in the sense of Peskun as fewer proposed moves will be accepted, but has a computational advantage as each component of the acceptance rate/event rate only requires evaluation of one pair-wise interaction.

To simulate a PDMP in practice it must be possible to either directly simulate from a Poisson process with intensity $\sum_{j\neq k}\lambda_{ij}(z)$, or to establish a tractable upper bound and then perform thinning. One approach to the latter is called the \emph{cell veto} method \citep{kapfer2016cell}, in which the manifold $\mathcal{M}$ is partitioned into cells and a local upper bound is found within each cell.  This approach has proved to be particularly effective for systems in which particles interact over long distances and has been applied by \cite{Faulkner2018AllAtomComputations} to the all-atom model of water presented in Section~\ref{sec:allatom}.  In this case the Metropolis algorithm requires prohibitively high per iteration costs, and molecular dynamics approaches are numerically unstable unless a very small step-size is chosen. Event chain Monte Carlo circumvents both issues, although work is still ongoing to improve molecular rotational mixing within the algorithm.  The cell-veto method has connections to a recently proposed approach for simulating PDMPs in the statistics literature by \cite{corbella2022automatic}.  

Alternative events are possible other than simply swapping the active particle using either the straight or reflected event chain strategies described above. \cite{michel2020forward} introduce \emph{forward} event chain Monte Carlo, in which the velocity $v$ is stochastically perturbed in a prescribed way when a collision event occurs. The motivation is that if enough randomness is introduced during this step then the algorithm can perform well even without introducing additional velocity refreshment events. \cite{Klementa2019EfficientEquilibration} and \cite{Hoellmer2022HardDiskDipoles} employ similar ideas in \emph{Newtonian} event chain Monte Carlo.  We add that parallel implementations of event chain Monte Carlo involving multiple active particles have also been considered in \cite{kampmann2015parallelized}.

\subsection{The Xtra chance algorithm}

The philosophy of continuing on the same path upon a rejection in the hope of reaching acceptance has also been proposed by \cite{campos2015extra} in the context of hybrid Monte Carlo.  In this algorithm Hamiltonian dynamics are numerically simulated for a period of time $T:=\ell \varepsilon$ and then an accept-reject decision is taken by sampling $u \sim \mathcal{U}[0,1]$ and assessing whether or not $u \leq \exp [- \beta (H(x',p') - H(x,p))]$, where $(x',p'):= \psi^\varepsilon_{\ell}(x,p)$ is the Hamiltonian proposal.  Upon rejection, however, a second-stage proposal is computed as $(x'',p''):= \psi^\varepsilon_\ell(x',p')$.  In other words, the dynamics are simulated for an additional time $T = \ell\varepsilon$.  This second stage proposal is accepted if $u < \exp [- \beta (H(x'',p'') - H(x,p))]$, where $u$ is the same uniform random variable used in the first stage accept-reject decision. The scheme has the appealing interpretation that first a $u$ is simulated, and then proposals are repeatedly tested until one is encountered for which the Metropolis ratio is larger than $u$. Typically a maximum number of attempts are tried until this goal is attained, otherwise all proposals are rejected.

The Xtra chance algorithm was designed so that the expensive computations associated with simulating Hamiltonian dynamics are re-used in the case of a rejection, as the end point of this simulation is used as the starting point for the next stage proposal. As in the case of Jaster's algorithm the probability of rejection is strictly decreased compared to ordinary hybrid Monte Carlo, but similarly more computation is associated with each iteration. The Xtra Chance algorithm can also be regarded as a particular case of the sequential proposals algorithm of \cite{park2020markov}, as is remarked in that work. In addition it can again be fitted into the delayed rejection framework of \cite{mira2001metropolis}. One surprising feature of the Xtra chance algorithm when viewed through the lens of delayed rejection is its attractively simple acceptance rate at the second stage (and beyond). In contrast, the usual acceptance rate for a delayed rejection algorithm is
$$
\alpha_2(x,x',x'') = \min\left(1, \frac{\pi(x'')q_1(x'',x')q_2(x'',x',x)(1-\alpha_1(x'',x'))}{\pi(x)q_1(x,x')q_2(x,x',x'')(1-\alpha_1(x,x'))}
\right),
$$
where $q_1(x,\cdot)$ and $q_2(x,x',\cdot)$ are the first and second stage proposal kernels and $\alpha_1(x,x')$ is the usual Metropolis--Hastings acceptance probability for the first stage proposal $x'$.  In addition, a fresh uniform random variable must be drawn to decide whether or not to accept the second stage proposal as compared to that used in the first stage.  The reason that a much simpler algorithm can be used in both the Xtra chance and Jaster algorithm is in part owing to the symmetries of the dynamics of the transition, but also to the augmented \textit{slice sampler} target density $\mu(x,u) := \mathbb{I}(u<\pi(x))$~\citep{neal2003slice}.  When viewing delayed rejection with $\mu$ as the target distribution, the acceptance rates for these algorithms reduce to being either 1 or 0, and the persistent uniform sample that determines when a proposal is accepted is nothing more than a transformed sample from the conditional distribution of $u|x$. \cite{andrieu2021peskun} prove that taking extra chances in this manner reduces the asymptotic variance of ergodic averages (the result does not, however, account for computational cost).


\subsection{Shadow hybrid Monte Carlo}

\cite{izaguirre2004shadow} introduced a modification to the hybrid Monte Carlo method that was later developed and introduced to the statistics community by \cite{radivojevic2020modified}. The shadow hybrid Monte Carlo method is motivated by the field of backward error analysis for ordinary differential equations (ODEs). Given an ODE $\dot{x}(t) = f(x(t))$ and a numerical scheme for simulating this ODE, \emph{forward} error analysis is concerned with understanding how far apart the numerical and exact solutions of the ODE are at some time step $t$, usually as a function of the numerical step-size $\varepsilon>0$. A scheme is called $p$th order accurate if this global error can be bounded by some function $C(t)\varepsilon^p$. The idea of \emph{backward} error analysis is to instead seek a \emph{modified} ODE system $\dot{x}(t) = \tilde{f}_\varepsilon(x(t))$ for which the numerical scheme that is used for the original ODE will be accurate to a higher order. The differences between $\tilde{f}_\varepsilon$ and $f$ can then be studied to understand qualitative differences in behaviour between the numerical scheme and the true solution to the original ODE. As a simple example, consider numerically simulating the system $\dot{x}(t) = f(x(t))$ using a first order scheme. Under sufficient smoothness assumptions on $f$, finding a modified system $\dot{x}(t) = \tilde{f}_\varepsilon(x(t))$ for which the scheme is second order accurate involves setting $\tilde{f}_\varepsilon(x) = f(x) + \varepsilon f_1(x)$ and then considering a Taylor series expansion of the one step error $\tilde{x}(t+\varepsilon) - \tilde{x}(t)$ to choose an $f_1$ that results in a cancellation of the relevant lower order terms in $\varepsilon$ (here $\tilde{x}(t)$ denotes the true solution of the modified system).  See e.g. Chapter 5 of \cite{leimkuhler2004simulating} for a more detailed explanation and examples.

Backward error analysis has found success in the study of symplectic numerical schemes for Hamiltonian systems, such as the velocity Verlet approach introduced in Section \ref{sec:ClassicalSamplingAlgorithmsMD}. Using this approach it is possible to show that in many cases the modified system is also Hamiltonian, and that the numerical scheme preserves the value of the modified or \emph{shadow} Hamiltonian over surprisingly long time scales \cite[Chapter~5]{leimkuhler2004simulating}. This gives some insight into the success of the hybrid Monte Carlo method.  It also presents opportunities to develop alternative approaches. The shadow hybrid Monte Carlo method works by simulating a hybrid Monte Carlo algorithm, but in place of the true Hamiltonian $H$ in the acceptance rate a chosen shadow Hamiltonian $\tilde{H}_k$ of order $k$ is used (the choice of order usually depends on trading off accuracy with computational cost). The shadow Hamiltonian $\tilde{H}_k$ will typically also depend on the momentum in a less straightforward way than $H$, meaning that in place of re-sampling the momentum directly from its marginal distribution a proposed change in momentum is drawn (from its marginal distribution in the true Hamiltonian system) and then accepted or rejected. \cite{izaguirre2004shadow} propose to use rejection sampling, whereas \cite{radivojevic2020modified} introduce a Metropolis step.  Rather than simple Monte Carlo averages of the algorithm output, which would give expectations with respect to the modified distribution with density $\propto e^{-\beta \tilde{H}_k}$, importance sampling estimators using weights $e^{- \beta (H - \tilde{H}_k)}$ can be employed, allowing expectations with respect to the true distribution of interest to be computed. \cite{radivojevic2020modified} consider several further modifications to the original algorithm, including different numerical integrators as introduced in \cite{radivojevic2018multi} and partial momentum refreshment as proposed in ordinary hybrid Monte Carlo by \cite{horowitz1991generalized}.

\begin{remark}
Combining importance sampling and Markov chain Monte Carlo in this way has also been proposed and studied in the statistics literature (e.g.~\cite{franks2020importance,vihola2020importance}). The intriguing property of the above scheme is the level of stability provided to the importance weights by the modified Hamiltonian. Traditional importance sampling typically performs poorly in high-dimensional settings, but there is much numerical evidence that this is not true of shadow hybrid Monte Carlo.
\end{remark}

\subsection{Sampling with modified kinetic energies}

The underdamped Langevin diffusion \eqref{eq:udlangevin} can be generalized in such a way that the invariant density for momentum is changed while that for the position variable $x$ remains $\pi(x) \propto e^{-\beta U(x)}$.  The resulting system of stochastic differential equations can be written analogously to \eqref{eq:udlangevin} as
\begin{equation} \label{eq:udlangevin_ar}
\begin{aligned}
dx_i(t) &= \nabla_i K(p(t))dt \\
dp_i(t) &= F_i(x(t))dt - \gamma \nabla_i K(p(t))dt + \sqrt{2\gamma \beta^{-1}}dW_i(t).
\end{aligned}
\end{equation}
where $K$ is some \emph{kinetic energy} function. The choice $K(p) = p^T M^{-1} p/2$ leads to \eqref{eq:udlangevin}, where $M$ is a diagonal matrix with entries $M_{ii} = m_i$.  
\cite{artemova2012adaptively} propose to modify the standard quadratic form choice of $K$ in such a way that a particle does not move if it has momentum below a chosen threshold $p_\textup{min}>0$, with the aim of reducing computational cost by freezing particles in place unless they will move by an appreciable amount. This is achieved by setting $K(p) := \sum_i k(p_i)$ where
$$
k(p_i) := 
\begin{cases}
0 & \|p_i\|_2 < p_\textup{min} \\
\frac{1}{2m_i}p_i^Tp_i & \|p_i\|_2 > p_\textup{max}.
\end{cases}
$$
In between the constants $p_\textup{min}$ and $p_\textup{max}$ the function smoothly interpolates between $0$ and $p_\textup{max}^2/(2m_i)$ \cite[Section~4.1.1]{stoltz2018langevin}. Ergodicity properties of \eqref{eq:udlangevin_ar} are studied in \cite{redon2016error} and guidelines for the choice of step-size as compared to the standard choice of kinetic energy are given in \cite{stoltz2018langevin}. A parallel implementation is introduced in \cite{singh2017parallel}.

Experiments in \cite{artemova2012adaptively} indicate that the resulting \emph{adaptively restrained} Langevin dynamics induce stronger autocorrelations over time among the particle positions, which is intuitive given that their movement is restricted. This is more than offset, however, by the seven-fold computational speed up exhibited by the restrained dynamics when compared to the standard underdamped Langevin approach, resulting in around four times better accuracy overall when comparing the error in estimating some chosen test functions in an example simulation of 343 Argon particles in which intra-molecular interactions were modelled using a Lennard--Jones potential. \cite{stoltz2018langevin} also consider the problem of metastability when sampling from multi-modal distributions, and provide numerical evidence that this can be reduced through an appropriate choice of kinetic energy.

\begin{remark}
The closely related idea of modifying the choice of kinetic energy within hybrid Monte Carlo has been considered in the statistics and machine learning literature in \cite{lu2017relativistic,zhang2016towards} and studied in terms of ergodicity properties in \cite{livingstone2019kinetic}. The main focus of these works, however, is on the increased robustness and numerical stability that can be achieved by using a slower growing kinetic energy than the standard choice. \cite{nishimura2020discontinuous} also consider different choices of kinetic energy in order to sample from discrete distributions and those with discontinuous potentials.
\end{remark}

\section{Simulation studies}
\label{sec:SimulationsStudies}

In this section we present simulation studies of the two-dimensional Ising and XY models, comparing the Metropolis algorithm with two of the advanced algorithms presented in Section~\ref{sec:AdvancedAlgorithms}.  For storage reasons, only an $N$-skeleton of the Metropolis chain is retained, corresponding to storing the system state after each $N$-particle sweep.  The elapsed simulation time during each $N$-particle sweep defines one unit of Metropolis simulation time $\Delta t_{\rm Metrop}$.  
The elapsed simulation time during each Wolff iteration defines one unit of Wolff simulation time $\Delta t_{\rm Wolff}$.

\subsection{Two-dimensional Ising model}
\label{sec:IsingSims}

\begin{figure*}
\includegraphics[width=\linewidth]{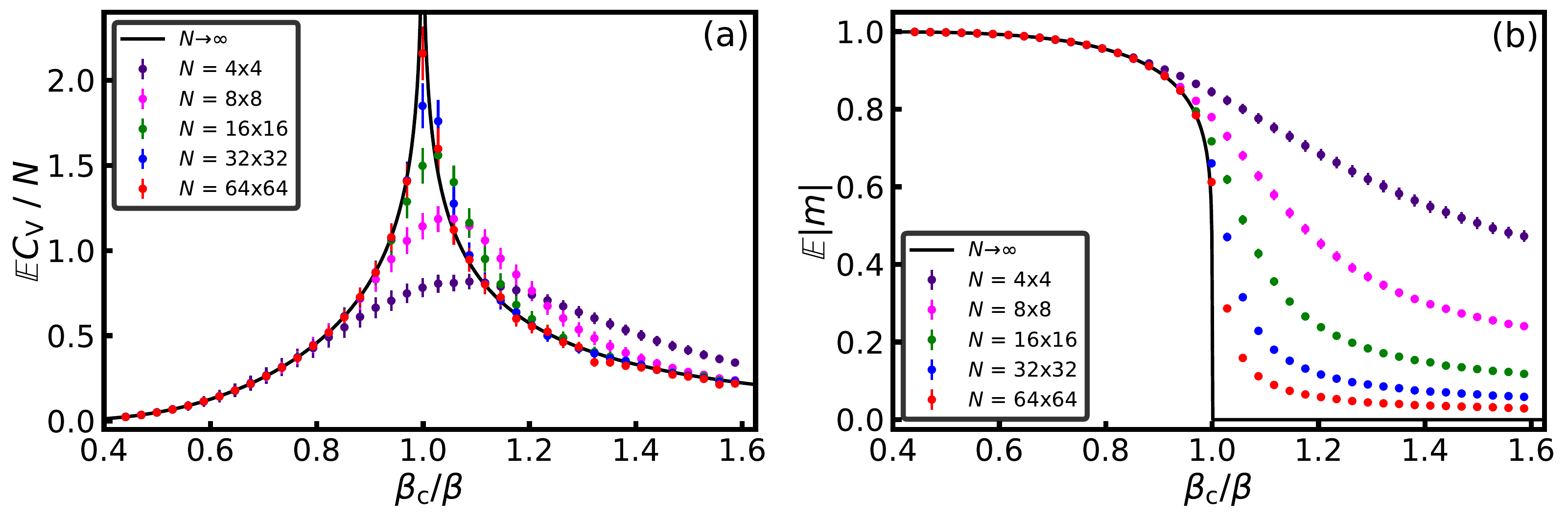}
\caption{Estimations of the expected specific heat per particle (a) and expected absolute magnetic density (b) of the two-dimensional Ising model as functions of $\beta_{\rm c} / \beta$ and number of particles $N$ at $h = 0$.  Results were generated from $10^4$ Wolff samples ($10^4$ burn-in iterations were discarded) and then averaged over $28$ simulations.  
The black curves in (a) and (b) correspond to the analytical thermodynamic predictions in \eqref{eq:2dIsingExpectedSpecHeat} and \eqref{eq:2dOnsagerYangSolution}.}
\label{fig:2dIsingSpecHeatAndAbsMagDensitySimulations}
\end{figure*}

Figure \ref{fig:2dIsingSpecHeatAndAbsMagDensitySimulations} shows estimates of the expected zero-field ($h = 0$) specific heat per particle (see \eqref{eq:ExpectedSpecificHeat}) and expected zero-field absolute magnetic density (see \eqref{eq:MagneticDensity}) of the two-dimensional Ising model, as functions of $\beta_{\rm c} / \beta$ and $N$ (recall that $\beta_{\rm c} := \ln (1 + \sqrt{2}) / (2J)$ is the inverse critical temperature). 
The output was generated using the Wolff algorithm 
and provides evidence for the phase transition predicted by \cite{Onsager1944CrystalStatistics} at $\beta = \beta_{\rm c}, h = 0$.  
The specific-heat output in Figure \ref{fig:2dIsingSpecHeatAndAbsMagDensitySimulations}(a) 
appears to approach the analytical thermodynamic prediction in \eqref{eq:2dIsingExpectedSpecHeat} with increasing $N$.  
The magnetic-density output in Figure \ref{fig:2dIsingSpecHeatAndAbsMagDensitySimulations}(b) suggests that the expected zero-field absolute magnetic density tends to the spontaneous magnetic density $m_0(\beta J)$ defined in \eqref{eq:2dOnsagerYangSolution} and corresponding to the solid black line in the figure. 
The output tends to one for all $N$ as $\beta \to \infty$ because the Boltzmann--Gibbs distribution puts all probability mass on two equally likely states ($x_i = 1$ for all $i$ and $x_i = - 1$ for all $i$) in this limit, while it appears to tend to zero in the thermodynamic limit for all $\beta < \beta_{\rm c}$ because the magnetic density $m$ 
satisfies $\mathbb{P}\left[ m (x; \beta, J, h = 0, N) \ne 0 \right] \to 0$ as $N \to \infty$ for all $\beta < \beta_{\rm c}$.  
Both outputs become noisier as $\beta$ becomes smaller and $N$ larger, which we discuss below.


\begin{figure*}
\includegraphics[width=\linewidth]{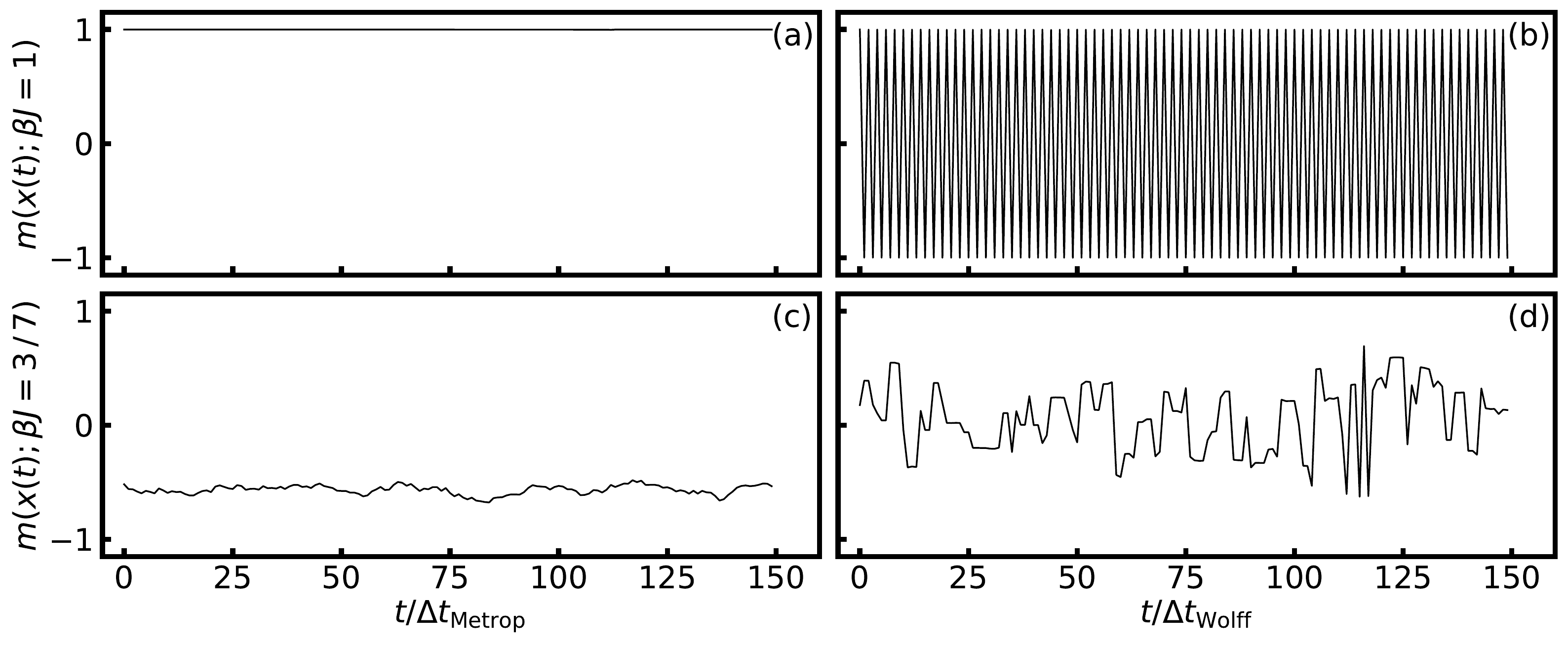}
\caption{Zero-field magnetic density $m(x(t); \beta, J, h = 0, N)$ versus normalised simulation time $t / \Delta t_{\rm Metrop/Wolff}$ for $N = 64 \! \times \! 64$ particles at low temperature ($\beta J = 1 > \beta_{\rm c} J$) and near $\beta_{\rm c}$ ($\beta J = 3 / 7$) using both the Metropolis (a, c) and Wolff (b, d) algorithms.  In each case $10^4$ burn-in iterations were discarded.}  
\label{fig:2dIsingMagDensityVsTime}
\end{figure*}

We now compare the Metropolis and Wolff algorithms in the context of \emph{spontaneous symmetry breaking}.  At $h = 0$ the potential is symmetric in $x$ for all $\beta, J, N$, but numerical simulations (of the system) constrained to single spin flips spontaneously break this $Z_2$ symmetry at finite $\beta > \beta_{\rm c}$, leaving the system stuck close to one of the two $\beta \to \infty$ states on a timescale that diverges with $N$.  This is an example of spontaneous symmetry breaking 
and is reflected in the low-temperature ($\beta > \beta_{\rm c}$) zero-field magnetic-density trace plots in Figures~\ref{fig:2dIsingMagDensityVsTime}(a) and (b).  On the presented simulation timescale, the Metropolis simulation starts at $m = 1$ and stays in this state, while the Wolff simulation mixes between $m = 1$ and $m = -1$.
This case of symmetry breaking is caused by the multi-modality of the Boltzmann--Gibbs distribution, but anisotropic distributions can also result in the phenomenon (e.g. the asymmetric Metropolis simulation output in Section~\ref{sec:XYSims}).

\begin{figure}
\includegraphics[width=0.5\linewidth]{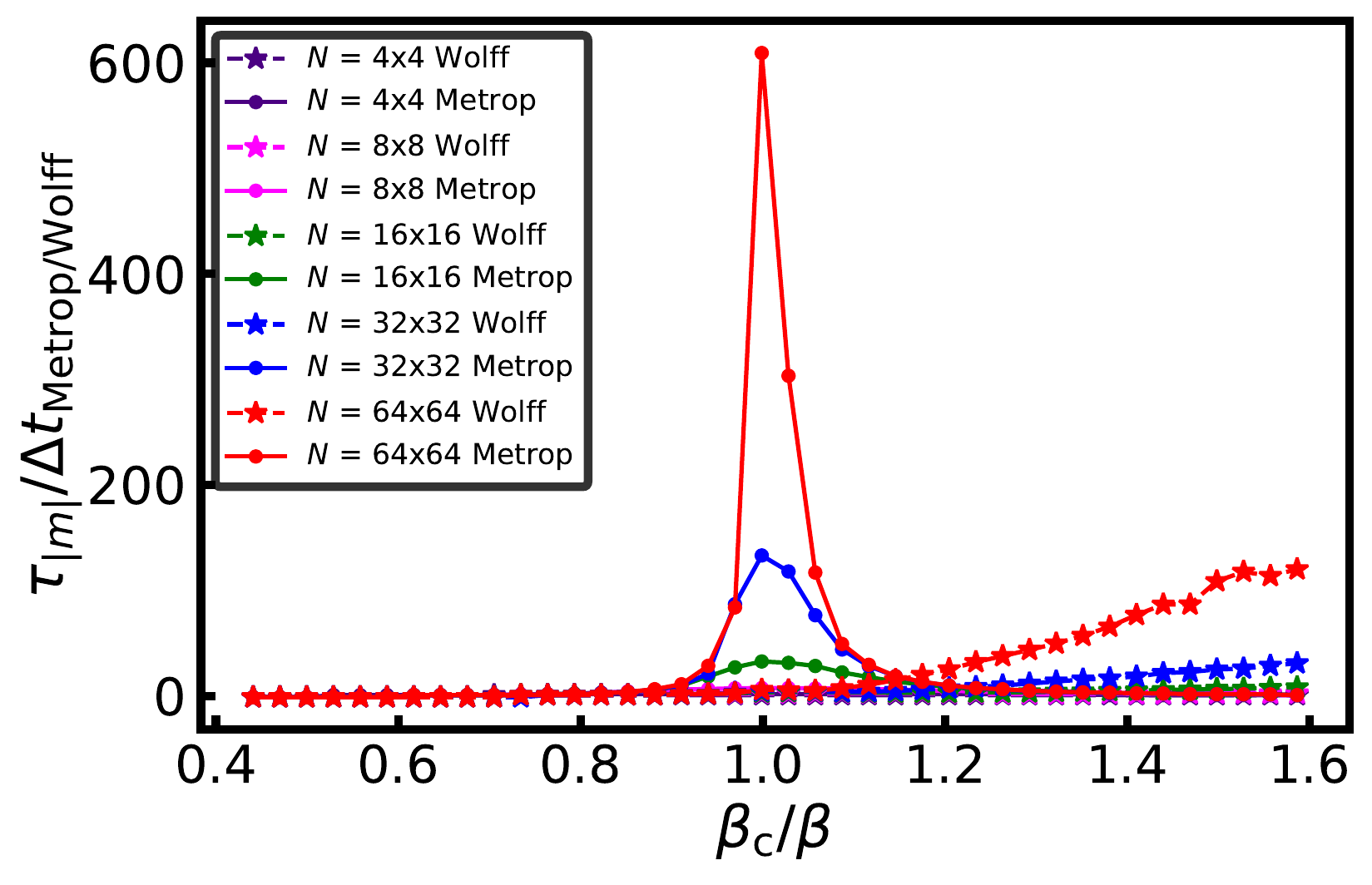}
\caption{Estimations of the normalised magnetic-norm integrated autocorrelation time $\tau_{|m|} / \Delta t_{\rm Metrop/Wolff}$ of the two-dimensional Ising model as a function of $\beta_{\rm c} / \beta$ and number of particles $N$, with respect to the Metropolis and Wolff algorithms.  Autocorrelation functions were generated from $10^4$ samples ($10^4$ burn-in iterations were discarded) and then averaged over $28$ simulations.}
\label{fig:2dIsingACF}
\end{figure}

The Wolff algorithm also combats the related phenomenon of \emph{critical slowing down}.  As $\beta_{\rm c}$ is approached from small values of $\beta$, 
strong particle--particle correlations begin to set in on increasingly long 
lattice-site-separation 
distances.  Near $\beta_{\rm c}$, this results in increasingly large clusters of particles with the same spin value, which slows mixing significantly for simulations constrained to single spin flips, resulting in very noisy $\check{m}$ statistics.  
This is remedied by flipping large clusters of spins, as reflected in the output in Figures~\ref{fig:2dIsingMagDensityVsTime}(c-d) and \ref{fig:2dIsingACF}.  Figures~\ref{fig:2dIsingMagDensityVsTime}(c) and (d) show (respectively) trace plots of the magnetic density at small $|\beta - \beta_{\rm c}|$ using the Metropolis and Wolff algorithms.  The Metropolis output is strongly time correlated, while the Wolff output mixes on the presented simulation timescale.  
Figure~\ref{fig:2dIsingACF} shows estimates of the magnetic-norm integrated autocorrelation times $\tau_{|m|}$ of the zero-field two-dimensional Ising model as functions of $\beta_{\rm c} / \beta$ and $N$, using both the Metropolis and Wolff algorithms.  
The Metropolis estimates appear to diverge with system size as $\beta \to \beta_{\rm c}$, while those of the Wolff algorithm depend only weakly on $N$ for all $\beta_{\rm c} / \beta < 1.2$, before developing a stronger $N$-dependence at smaller $\beta$, due to smaller typical cluster sizes.  This results in the development of noise in the output in Figure~\ref{fig:2dIsingSpecHeatAndAbsMagDensitySimulations} at small $\beta$.  Physicists tend to remove any $N$-dependence from the Wolff timescale by multiplying it by some metric for the typical cluster density at any given temperature (e.g.~\cite{Tamayo1990SingleClusterMC}). 

\subsection{Two-dimensional XY model}
\label{sec:XYSims}

\begin{figure*}
\includegraphics[width=\linewidth]{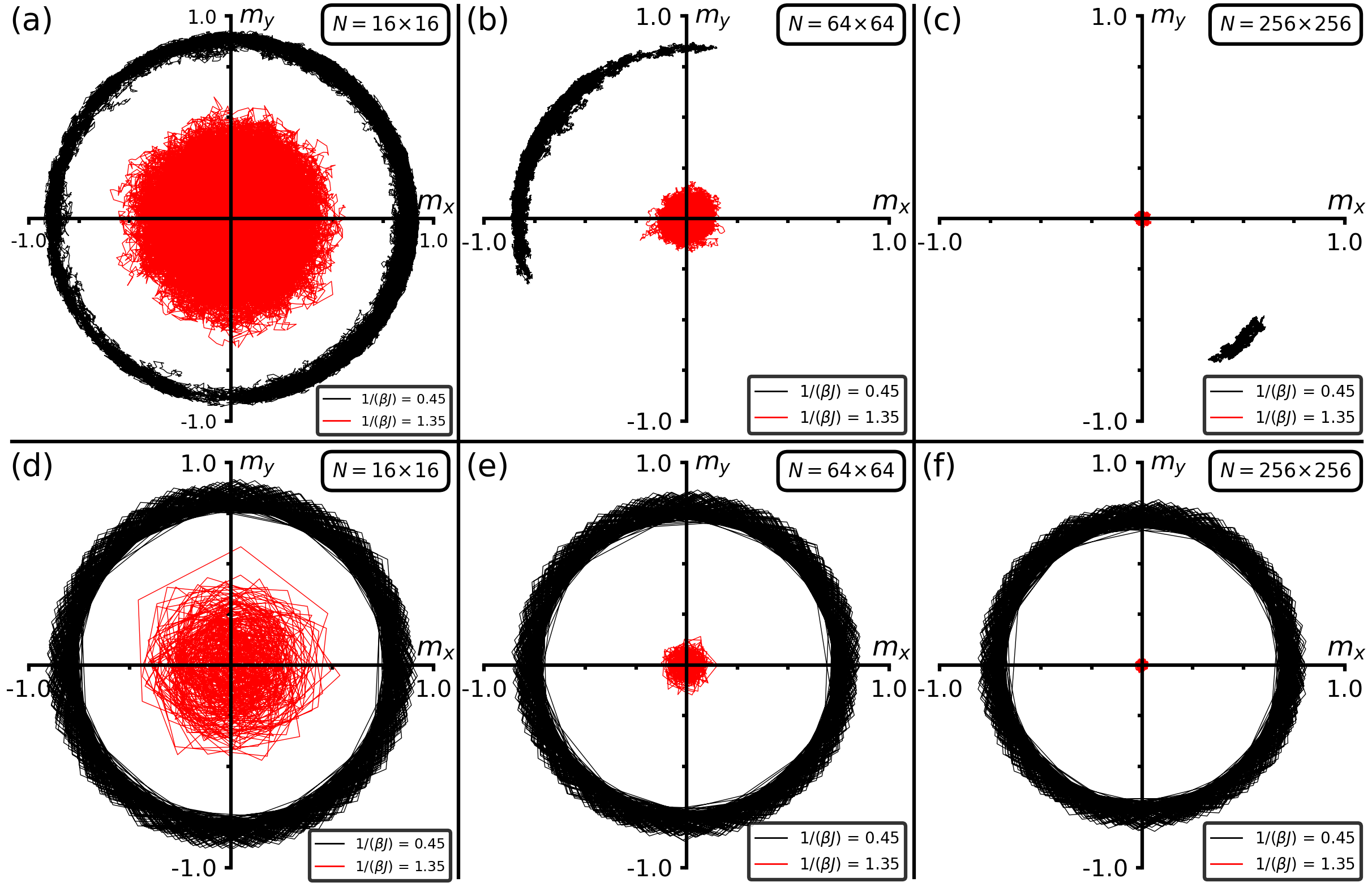}
\caption{Evolution of the magnetic density $m$ of the two-dimensional XY model 
using the Metropolis (a-c) and event-chain (d-f) algorithms. 
The size of the Metropolis skeleton chain is $10^5$ (with acceptance probability $\simeq 0.6$) while that of the event-chain algorithm is $10^3$.}
\label{fig:XYMagnetisationEvolution}
\end{figure*}

In addition to its use in the development of the theory of the two-dimensional melting transition presented in Section \ref{sec:harddisk}, event chain Monte Carlo applied to the XY model has also enjoyed success analogous to that of the Wolff algorithm for the Ising model.  Figure~\ref{fig:XYMagnetisationEvolution} shows the evolution of the two-dimensional magnetic density vector (as defined in \eqref{eq:MagneticDensity}) of the zero-field two-dimensional XY model (presented in Section \ref{sec:XYandPotts}) using both the Metropolis and event chain Monte Carlo algorithms.  The expected value is $(0,0)$ at all nonzero temperatures, but the low-temperature Metropolis mean converges to this expected value on a timescale that diverges with $N$, as described in detail in  \cite{Faulkner2022GeneralSymmetryBreaking} and reflected in Figures~\ref{fig:XYMagnetisationEvolution}(a-c) (`low temperature' corresponds to all finite $\beta > \beta_{\rm BKT}$, where $\beta_{\rm BKT} \simeq 1.13 / J$ is the inverse critical temperature of the model).
The event-chain output (Figures~\ref{fig:XYMagnetisationEvolution}(d-f)) by contrast suggests $N$-independence for all $\beta$.  The event chain algorithm was simulated for a time period proportional to $N$.


\section{Discussion}
\label{sec:Discussion}

It is important to acknowledge that there has been much successful cross-pollination of ideas between statistical physics and statistics/machine learning for many decades now. The perhaps unfortunately named \textit{Gibbs sampler} is so-called from its initial use in sampling from Ising models applied to pixel distributions in black and white images \citep{geman1984stochastic}. The now well-established use of energy-based models in image processing \citep{lecun2006tutorial} is an example of how such cross-pollination can lead to new approaches and insights. Nonetheless, we believe that there is still much that each field can learn from the other, and speculate on some possible avenues below.

The ergodicity properties of the most successful version of (generalized) event chain Monte Carlo are, to the best knowledge of the authors, still largely unknown (see Remark \ref{rem:ergodicityECMC} of Section \ref{sec:ecmc}). Statisticians have recently found success in studying ergodicity properties of piecewise deterministic Markov processes (e.g. \cite{bierkens2019ergodicity}) and may have something to offer here. In addition, there is a need to establish theoretical results comparing the straight and reflected versions of event chain Monte Carlo, as well as uniform versus $xy$ refreshment of particle velocities, to support empirical findings. Statisticians have some track record in establishing orderings among sampling algorithms, and the accumulated knowledge within the field could be suitable for this task.

Often when hybrid (or \textit{Hamiltonian}) Monte Carlo is described in statistics and machine learning, an analogy is given to the movement of a single $Nd$-dimensional fictitious `particle' (e.g. \cite{neal2011mcmc}). Physicists, by contrast, think separately about the movement of each individual $d$-dimensional \textit{physical} particle within a system. With this mindset interactions between particles become a natural consideration, and it may be that such a mindset is beneficial in the context of probabilistic models. One example is sampling under ordering constraints, such as learning the positions of histogram breaks \citep{worrall2022fifty} or knot positions in a nonparametric regression model \citep{smith1996nonparametric}. One-dimensional implementations of molecular or event chain dynamics would trivially preserve the necessary monotonicity requirements imposed by such models.  Repulsive potentials such as Lennard--Jones could also be incorporated into prior distributions for parameters in a Bayesian setting to ensure that sufficiently different posterior distributions are attained, such as in the estimation of parameters in mixture models. 

In the field of dimension reduction ideas from statistical physics are implicitly used.  In multi-dimensional scaling \citep{torgerson1952multidimensional} $d$-dimensional representations $x_i$ for $1 \leq i \leq N$ of some higher dimensional data points $y_1,...,y_N$ are sought in order to reduce the dimension of the problem. In metric multi-dimensional scaling, estimates for $x_1,...,x_N$ are found by minimising the \emph{stress} function $S(x_1,...,x_N) := \sum_{i\neq j} (d_{ij} - \metric(x_i,x_j))^2$, where $d_{ij}$ is some appropriate notion of dissimilarity between $y_i$ and $y_j$ computed in the original high-dimensional space.\footnote{Often the square root of $S$ is defined as the stress, but this operation has no impact on the minima.} The stress function, when viewed through the lens of statistical physics, is nothing more than a bonded potential. This observation has in fact already been noted \citep{andrecut2009molecular}, but there is still much scope to build on the connection. In particular there is recent interest in quantifying uncertainty in the parameters $x_1,...,x_N$ through a Bayesian treatment (e.g. \cite{ren2017bayesian}), and viewing the problem through the lens of Boltzmann--Gibbs distributions, in particular when combined with recent advances in model-free Bayesian inference \citep{bissiri2016general,jewson2018principles,knoblauch2022optimization}, may be a fruitful avenue for this problem.

This article has emphasised that the study of phase transitions is central to statistical physics. In a typical problem, samples are drawn from several different Boltzmann--Gibbs distributions, each corresponding to a different fixed temperature (or some other fixed hyperparameter of the system), and the qualitative behaviour of some observable is studied as a function of temperature. One example of a `phase transition' in statistics/machine learning is the estimator for a coefficient value in L1-penalised regression, as a function of the weight $\lambda$. The inverse temperature parameter is also present (albeit with the addition of a prior) as the \textit{learning rate} in Gibbs posterior distributions (e.g. \cite{haddouche2021pac,bissiri2016general,syring2020gibbs,grunwald2017inconsistency}). Typically the goal has been to choose a unique optimal value for this parameter \citep{wu2022comparison}, but summary statistics of interest could also be treated as `observables' and studied for evidence of phase transitions as the learning rate varies.

There are numerous approaches to sampling from Boltzmann--Gibbs distributions that were omitted for brevity. Nothing is mentioned of thermostats, such as Nos\'{e}--Hoover dynamics and extensions (e.g. \cite{evans1985nose,martyna1992nose,leimkuhler2009gentle}). Details of this approach and a simulation study comparing to hybrid Monte Carlo are provided in \cite{cances2007theoretical}. Biasing methods such as the Wang--Landau algorithm \citep{Wang2001Efficient} and the adaptive biasing force method \citep{darve2008adaptive} were also not discussed. We refer the interested reader to book-level treatments of the subject such as \cite{lelievre2010free,leimkuhler2016molecular,krauth2006statistical}.

We have also omitted the estimation of dynamical properties.  
Barrier-crossing techniques such as metadynamics \citep{barducci2011metadynamics} or multicanonical sampling \citep{berg1991multicanonical} can be integrated into sampling algorithms to estimate the depths and shapes of potential wells, providing a route to order-of-magnitude estimates of chemical-reaction rates, rare-event timescales, and other similar phenomena.  More direct dynamical quantities may also be estimated when simulating steady-state systems (such as transport coefficients describing particle flow in response to an applied field).  These are typically the solution of a partial differential equation, though Metropolis simulations of such systems often produce remarkably accurate results (e.g.~\cite{Kaiser2013OnsagersWienEffect,Kaiser2015ACWienEffectInSpinIce}).

Finally, we have said very little here of algorithms that have been developed in statistics/machine learning but may have uses in statistical physics.  Recent advances in sampling on discrete state spaces such as the locally-informed approach of \cite{zanella2020informed} and the non-reversible strategies outlined in \cite{power2019accelerated} are a good example of this.  Several methods for estimating normalising constants have also been developed, such as Annealed Importance Sampling \citep{neal2001annealed}. These methods could well be useful in the context of free energy calculations. We have deliberately restricted ourselves to algorithms that originated from statistical physics, but a future article could certainly be written introducing these methods to an audience of physicists.




\vspace{1em}
\noindent
{\bf Code and data availability}.  All code used in this article is freely available on GitHub at \href{https://github.com/michaelfaulkner/super-aLby}{https://github.com/michaelfaulkner/super-aLby}, commit hash \href{https://github.com/michaelfaulkner/super-aLby/commit/1c014ca4536e3da3e7ed0dba307d97cb769b1574}{1c014ca} (Ising simulations) and \href{https://github.com/michaelfaulkner/xy-type-models}{https://github.com/michaelfaulkner/xy-type-models}, commit hash \href{https://github.com/michaelfaulkner/xy-type-models/commit/95093aabce2a31337c9959115348fcd82b192f1c}{95093aa} (XY simulations).  All published data can be reproduced using these applications (as outlined in each README) and are available at the University of Bristol data repository, \href{https://data.bris.ac.uk/data/}{data.bris}, at \href{https://doi.org/10.5523/bris.sju7uasr7e2b2n518hk72p3ur}{https://doi.org/10.5523/bris.sju7uasr7e2b2n518hk72p3ur}.

\vspace{1em}
\noindent
{\bf Acknowledgements}.  This work was conceived at the \href{https://sms.cam.ac.uk/collection/2519802}{Scalable inference; statistical, algorithmic, computational aspects} workshop (part of \href{http://www.i-like.org.uk}{i-like}, an EPSRC programme grant) at the Isaac Newton Institute for Mathematical Sciences, University of Cambridge.  All simulations were performed on BlueCrystal 4 at the Advanced Computing Research Centre, University of Bristol.  MFF acknowledges support from EPSRC fellowship EP/P033830/1.  SL acknowledges support from EPSRC grant EP/V055380/1.


%

%
%

\begin{supplement}
\stitle{Additional derivations and proofs}  
\sdescription{Free-energy derivation for the one-dimensional Ising model and proofs of Propositions \ref{prop:swendsenwang} and \ref{prop:ECqsa} can be found in the appendices of the present manuscript.}
\end{supplement}

\appendix
\section{Free-energy derivation for one-dimensional Ising model}
For a two-particle system, the partition function of the one-dimensional Ising model is 
\begin{align}
Z_{{\rm Ising, } d = 1}(\beta, J, h, N = 2) =& \sum_{x_1 = \pm 1} \sum_{x_2 = \pm 1} e^{\beta J x_1 x_2 + \frac{\beta h}{2} (x_1 + x_2)} e^{\beta J x_2 x_1 + \frac{\beta h}{2} (x_2 + x_1)}  \nonumber \\
=& e^{2 \beta J} \left( e^{2 \beta h} + e^{-2 \beta h} \right) + 2 e^{-2 \beta J} . \nonumber
\end{align}
Defining the matrix 
\begin{align}
P(\beta, J, h) := 
\begin{bmatrix}
e^{\beta (J + h)} & e^{- \beta J} \\
e^{- \beta J} & e^{\beta (J - h)} 
\end{bmatrix} , \nonumber
\end{align}
this can be rewritten as 
\begin{align}
Z_{{\rm Ising, } d = 1}(\beta, J, h, N = 2) = {\rm tr} \left[ P^2(\beta, J, h) \right] . \nonumber
\end{align}
For $N > 1$ particles, this then generalises to 
\begin{align}
Z_{{\rm Ising, } d = 1}(\beta, J, h, N) = {\rm tr} \left[ P^N(\beta, J, h) \right] , \nonumber
\end{align}
which follows from the factorisation of the partition function into the product of $N$ two-particle terms:
\begin{align}
Z_{{\rm Ising, } d = 1}(\beta, J, h, N) = \sum_{x_1 = \pm 1} \dots \sum_{x_N = \pm 1} e^{\beta J x_1 x_2 + \frac{\beta h}{2} (x_1 + x_2)} \dots e^{\beta J x_N x_1 + \frac{\beta h}{2} (x_N + x_1)} . \nonumber 
\end{align}
Since $P(\beta, J, h)$ is a symmetric, real-valued matrix, ${\rm tr} \glc P^N(\beta, J, h) \grc = \lambda_+^N + \lambda_-^N$, where 
\begin{align}
    \lambda_{\pm} (\beta, J, h) = e^{\beta J} \glc \coshb{\beta h} \pm \sqrt{\sinh^2(\beta h) + e^{-4\beta J}} \grc \nonumber
\end{align} 
are the two eigenvalues of $P(\beta, J, h)$.  It then follows that the free energy is 
\begin{align}
F_{{\rm Ising, } d = 1}(\beta, J, h, N) = - \beta^{-1} \logc{\lambda_+^N (\beta, J, h) + \lambda_-^N (\beta, J, h)} . 
\end{align}

\section{Miscellaneous proofs}

\begin{proof}[Proof of Proposition 2]
Consider the augmented state space $(x,b)$ and the joint distribution
\begin{equation}
    \mu(x,b) \propto e^{-\beta U_\textup{Ising}(x; J, 0, N)} \left( \prod_{i=1}^N \prod_{j \in S_i} q_{ij}(x)^{b_{ij}/2}(1-q_{ij}(x))^{(1-b_{ij})/2} \right).
\end{equation}
We first consider $\pi$-invariance, followed by irreducibility and aperiodicity.  On this augmented state space the Swendsen-Wang transition can be viewed as the combination of two updates applied sequentially. In the first we simply re-sample $b|x$ from its conditional distribution, which is clearly a $\mu$-preserving transition.  In the second we update $x|b$ by flipping the signs of spins within each cluster with probability 1/2.  The transition probability associated with this second step can be written
$$
P_\textup{SW}((x,b),(x',b')) = 2^{-C(b)} \mathbb{I}(b = b') \left( \prod_{i=1}^N \prod_{j \in S_i} b_{ij}\mathbb{I}(x_i' = x_j')\mathbb{I}(x_i = x_j) \right)^\frac{1}{2}.
$$
The first term on the right-hand side is a normalising constant, in which $C(b)$ denotes the number of clusters in the partition induced by $b$.  The second term stipulates that $b$ does not change. The third ensures that if a bond exists between particles $i$ and $j$ then they must take the same value. The final indicator function $\mathbb{I}(x_i = x_j)$ is not strictly necessary as provided $b$ is drawn from its conditional distribution given $x$ then $b_{ij}\mathbb{I}(x_i = x_j) = b_{ij}$, as a bond can only exist between particles $i$ and $j$ if they have the same spin. It does, however, make it clear that $P_\textup{SW}((x,b),(x',b')) = P_\textup{SW}((x',b'),(x,b))$, meaning that $\mu-$reversibility follows from showing that
\begin{equation} \label{eq:SWreversibility}
e^{-\beta \left( U_\textup{Ising}(x';J,0,N) - U_\textup{Ising}(x;J,0,N) \right) } = \prod_{i=1}^N \prod_{j \in S_i} \frac{q_{ij}(x)^{b_{ij}/2}(1-q_{ij}(x))^{(1-b_{ij})/2}}{q_{ij}(x')^{b_{ij}/2}(1-q_{ij}(x'))^{(1-b_{ij})/2}}.
\end{equation}
This can be seen by first considering the left-hand side of \eqref{eq:SWreversibility} and noting that $x_ix_j = 2\mathbb{I}(x_i = x_j) - 1$ when $x_i$ and $x_j$ can only take the values $\{-1,+1\}$, meaning
$$
\begin{aligned}
U_\textup{Ising}(x';J,0,N) - U_\textup{Ising}(x;J,0,N) &= -J \sum_{i=1}^N \sum_{j \in S_i} \left[ \mathbb{I}(x_i' = x_j') - \mathbb{I}(x_i = x_j) \right].
\end{aligned}
$$
This can be further modified by noting that under the Swendsen--Wang update the function $\mathbb{I}(x_i = x_j) - \mathbb{I}(x_i' = x_j')$ can only be non-zero for neighbouring particles $i$ and $j$ if $b_{ij} = 0$, meaning that if $x'$ is generated from such an update then
$$
\begin{aligned}
U_\textup{Ising}(x';J,0,N) - U_\textup{Ising}(x;J,0,N) &= -J \sum_{i=1}^N \sum_{j \in S_i} \left[ \mathbb{I}(x_i' = x_j') - \mathbb{I}(x_i = x_j) \right](1-b_{ij}).
\end{aligned}
$$
Turning to the right-hand side of \eqref{eq:SWreversibility} notice first that since $q_{ij}(x) = 1-e^{-2\beta J \mathbb{I}(x_i=x_j)}$ then $q_{ij}(x) = q_{ij}(x')$ when $b_{ij} = 1$, meaning that upon substituting in the definition of $q_{ij}(x)$ the fraction can be re-written
$$
\prod_{i=1}^N \prod_{j \in S_i} \frac{e^{-\beta J \mathbb{I}(x_i = x_j)(1-b_{ij})}}{e^{-\beta J \mathbb{I}(x_i' = x_j')(1-b_{ij})}},
$$
which from the calculations above is clearly equal to $e^{-\beta (U_\textup{Ising}(x';J,0,N) - U_\textup{Ising}(x;J,0,N)) }$ as required.

Establishing irreducibility and aperiodicity is straightforward. Aperiodicity can be seen by noting that the algorithm has a positive probability of not moving.  Irreducibility can be seen by simply noting that for any fixed $N$ there is a positive probability that $\prod_{i=1}^N \prod_{j \in S_i} b_{ij} = 0$ regardless of the value of the current state $x$. In this instance each particle belongs to its own cluster and therefore $x'$ can take any value in $\{-1,+1\}^N$ with the same (non-zero) probability. Any $x'$ can be arrived at from any $x$ in a single iteration of the algorithm. Similarly any $b'$ can be arrived at from any $b$ by first transitioning through an appropriate $x'$, since for every configuration of bonds $b'$ there is a configuration of spins $x'$ for which $b'|x'$ has non-zero probability.  The chain therefore has limiting distribution $\mu(x,b)$, and the marginal process on $x$ has limiting distribution $\pi(x) \propto e^{-\beta U_\textup{Ising}(x;J,0,n)}$, from which ergodicity follows.
\end{proof}

\begin{proof}[Proof of Proposition 3]
The result has already been established for the first two parts of the infinitesimal generator in Appendix E of \cite{andrieu2020general}. To extend to the $xy$-version, it suffices therefore to show $(\mu,Q)$-self-adjointness of the last part when $R$ implements the $xy$ transformation.

For the case $d=2$ note that $R(z,dz')$ simply swaps $(u_1,u_2) \to (u_2,u_1)$. We write $v \to \mathcal{S}(v)$ for this transformation and note that $\mathcal{S}(-v) = -\mathcal{S}(v)$ and $\mathcal{S}\circ \mathcal{S}(v) = v$, meaning $\mathcal{S}$ is an involution. Since $\langle f,g\rangle_\mu = \langle Qf,Qg\rangle_\mu$ and $Q^2 = I$ then setting
\begin{equation}
    L'f(x,v,i) := \lambda_\textup{ref} [f(x,\mathcal{S}(v),i) - f(x,v,i)]
\end{equation}
one can equivalently show that $\langle Q L'f,g\rangle_\mu = \langle f,QL'g\rangle_\mu$, meaning that the $Q$-symmetrization $QL'$ is $\mu$-self-adjoint (see \cite{andrieu2021peskun} for more detail). Direct calculation gives
\begin{equation}
\begin{aligned}
\lambda_{\rm ref}^{-1}\langle QL'f, g\rangle_\mu &= \int [f(x,-\mathcal{S}(v),i) - f(x,-v,i)]g(x,v,i) d\mu 
\\
&= \int f(x,-\mathcal{S}(v),i)g(x,v,i)d\mu - \int f(x,-v,i)g(x,v,i)d\mu.
\end{aligned}
\end{equation}
We can apply the change of variables $v \to -\mathcal{S}(v)$ to the first integral and $v \to -v$ to the second.  Since $\mu$ is invariant to both, the expression becomes
\begin{equation}
    \int [g(x,-\mathcal{S}(v),i) - g(x,-v,i)]f(x,v,i)d\mu = \lambda_{\rm ref}^{-1}\langle f,QL'g\rangle_\mu,
\end{equation}
which completes the proof.

\end{proof}

\begin{remark}
The $d>2$ setting can also be considered. In this case $\mathcal{S}$ is no longer an involution, but it is invertible, with $\mathcal{S}^{-1}(v)$ simply shifting each element of $u$ one space to the left modulo $d$.  Following Remark 5 of \cite{andrieu2020general} we can therefore introduce the auxiliary $w \in \{-1,+1\}$ and define the involution $\mathcal{S}(v,w) := (\mathcal{S}^w(v),-w)$ on an extended space. We can then incorporate $w$ into the augmented state $(x,v,i,w)$ and augment the measure $\mu$ to include a symmetric component for $w$, then perform analogous calculations to those above to establish the result. We omit the details for brevity.
\end{remark}


\bibliographystyle{imsart-nameyear} 
\bibliography{General,SamplingAlgorithms}       


\end{document}